\documentclass[10pt]{article}
\usepackage{ametsoc2col}

\usepackage{comment} 
\usepackage{graphicx}

\usepackage{microtype}
\usepackage[font=footnotesize,labelfont=bf]{caption}
\usepackage{amsmath,amsfonts,amssymb,bm}

\usepackage{ulem}

\usepackage{hyperref}
\hypersetup{
	hyperindex,
	breaklinks,
	colorlinks=true,
	linkcolor=blue,
	citecolor=blue,
	bookmarks=true,
	bookmarksopen=true,
	bookmarksopenlevel=2,
	pdfstartpage={1},
	pdfstartview={FitH},
	pdfview={FitH 0},
	pdfauthor={Navid C. Constantinou, Brian F. Farrell and Petros J. Ioannou},
	pdftitle={Emergence and equilibration of jets in beta-plane turbulence: applications of Stochastic Structural Stability Theory},
 }
\ifthenelse{\boolean{dc}}
{
\usepackage[all]{hypcap}	
}
{}

\usepackage{natbib}
\usepackage{doi}

\usepackage[usenames,dvipsnames]{xcolor}

\def\U{\bm{\mathsf{U}}}
\def\Uv{\mathbf{U}}

\def\C{\bm{\mathsf{C}}}
\def\I{\bm{\mathsf{I}}}

\def\D{{\bf D}}

\def\D{\bm{\mathsf{D}}}
\def\A{\bm{\mathsf{A}}}
\def\F{\bm{\mathsf{F}}}

\def\Q{\bm{\mathsf{Q}}}
\def\M{\bm{\mathsf{M}}}

\newcounter{saveeqn}%

\newcommand{\be}{\begin{equation}}
\newcommand{\ee}{\end{equation}}
\newcommand{\bdm}{\begin{equation*}}
\newcommand{\edm}{\end{equation*}}
\newcommand{\bea}{\begin{eqnarray}}
\newcommand{\eea}{\end{eqnarray}}

\newcommand{\partialf}[2]
{
 \ifthenelse{\equal{#1}{}}{\frac{\partial}{\partial #2}}{\frac{\partial #1}{\partial #2}}
}

\newcommand{\real}{\mathop{\mathrm{Re}}}
\newcommand{\imag}{\mathop{\mathrm{Im}}}
\newcommand{\vecd}{\mathop{\mathrm{vecd}}}

\newcommand{\Tr}{\mathop{\mathrm{Tr}}}
\renewcommand{\(}{\left(}
\renewcommand{\)}{\right)}
\renewcommand{\[}{\left[}
\renewcommand{\]}{\right]}
\newcommand{\<}{\left\langle}
\renewcommand{\>}{\right\rangle}

\newcommand{\Del}{\Delta}
\newcommand{\DDel}{\bm{\mathsf{\Del}}}
\renewcommand{\d}{\delta}

\newcommand{\df}{\textrm{d}}
\renewcommand{\l}{\ell}

\newcommand{\s}{\sigma}

\newcommand{\rU}{r_{\textrm{m}}}

\renewcommand{\v}{v}
\newcommand{\z}{\zeta}

\newcommand{\zv}{\mbox{\boldmath$\zeta$}}

\renewcommand{\i}{\mathrm{i}}

\newsavebox{\astrutbox}
\sbox{\astrutbox}{\rule[-5pt]{0pt}{20pt}}

\usepackage{upgreek}

\newcommand{\transp}{\textrm{T}}

\def\bit{\vphantom{\dot{W}}}

\newcommand{\myabstract}{
Stochastic Structural Stability Theory (S3T) provides analytical methods for understanding the emergence and equilibration of jets from the turbulence in planetary atmospheres based on the dynamics of the statistical mean state of the turbulence closed at second order. Predictions for formation and equilibration of turbulent jets made using S3T are critically compared with results of simulations made using the associated quasi-linear and nonlinear models. S3T predicts the observed bifurcation behavior associated with the emergence of jets, their equilibration and their breakdown as a function of parameters. Quantitative differences in bifurcation parameter values between predictions of S3T and results of nonlinear simulations are traced to modification of the eddy spectrum which results from two processes: nonlinear eddy-eddy interactions and formation of discrete non-zonal structures. Remarkably, these non-zonal structures, which substantially modify the turbulence spectrum, are found to arise from S3T instability. Formation as linear instabilities and equilibration at finite amplitude of multiple equilibria for identical parameter values in the form of jets with distinct meridional wavenumbers  is verified as is the existence at equilibrium of finite amplitude non-zonal structures in the form of nonlinearly modified Rossby waves. When zonal jets and nonlinearly modified Rossby waves coexist at finite amplitude the jet structure is generally found to dominate even if it is linearly less unstable. The physical reality of the manifold of S3T jets and non-zonal structures is underscored by the existence in nonlinear simulations of jet structure at subcritical S3T parameter values which are identified with stable S3T jet modes  excited by turbulent fluctuations.
}

\begin{document}

\title{\textbf{\large{Emergence and equilibration of jets in beta-plane turbulence:\\applications of Stochastic Structural Stability Theory}}}

\author{\textsc{Navid C. Constantinou}\thanks{\textit{Corresponding author address:} Navid Constantinou, University of Athens, Department of Physics, Section of Astrophysics, Astronomy and Mechanics, Build IV, Office 32, Panepistimiopolis, 15784 Zografos, Athens, Greece.\newline{E-mail: \href{mailto:navidcon@phys.uoa.gr}{navidcon@phys.uoa.gr}}}\\
\textit{\footnotesize{Department of Physics, National and Kapodistrian University of Athens, Athens, Greece}}
\and
\centerline{\textsc{Brian F. Farrell}}\\
\centerline{\textit{\footnotesize{Department of Earth and Planetary Sciences, Harvard University, Cambridge, MA 02138}}}\\
\and
\textsc{Petros J. Ioannou}\\
\textit{\footnotesize{Department of Physics, National and Kapodistrian University of Athens, Athens, Greece}}
\\
\\
\centerline{\small (Submitted to the Journal of the Atmospheric Sciences)}
}

\ifthenelse{\boolean{dc}}
{
\twocolumn[
\begin{@twocolumnfalse}
\amstitle

\begin{center}
\begin{minipage}{13.0cm}
\begin{abstract}
	\myabstract
	\newline
	\begin{center}
		\rule{38mm}{0.2mm}
	\end{center}
\end{abstract}

\end{minipage}
\end{center}
\end{@twocolumnfalse}
]
}
{
\amstitle
\begin{abstract}
\myabstract
\end{abstract}
\newpage
}



\section{Introduction}

Spatially and temporally coherent jets are a common feature of turbulent flows in planetary atmospheres  with the banded winds of the giant planets constituting a familiar  example  
\citep{Vasavada-and-Showman-05}. \cite{Fjortoft-1953} noted that the conservation of both energy and enstrophy in dissipationless barotropic flow implies that transfer of energy among spatial spectral components results in energy 
accumulating at the largest  scales. This argument provides a conceptual basis for understanding the 
observed tendency for formation  of large scale structure from small scale turbulence in  planetary atmospheres. 
However, the observed large scale structure is dominated by zonal jets with specific form and, moreover,  the scale of these jets is distinct from the largest scale  in the flow. 
\cite{Rhines-1975}  argued that  the observed spatial scale of jets in beta-plane turbulence results from arrest of 
upscale energy transport
at the  length scale, $\sqrt{u/\beta}$, where $\beta$ is the meridional gradient of planetary vorticity and $u$ is the root mean square velocity in the turbulent fluid. 
In Rhines' interpretation this is the scale at which the turbulent energy cascade is intercepted 
by the formation of propagating Rossby waves. 
\cite{Balk-etal-1991} extended Rhines' argument by showing  that in addition to energy and enstrophy, dissipationless barotropic turbulence conserves a  third quadratic  invariant, called zonostrophy, which constrains the large scale structures in dissipationless beta plane turbulence to be predominantly zonal (cf.~\cite{Balk-etal-2009}).
While these results establish a
conceptual basis for expecting large scale zonal structures to form in beta plane turbulence, the physical mechanism of jet formation, the structure of the jets, and their dependence on parameters remain to be determined. 

One mechanism for formation of jets is  vorticity  mixing resulting from  
Rossby wave breaking which leads to homogenization of vorticity in localized regions and formation of
vorticity  staircases. The risers of these staircases correspond  to thin  prograde  jets located at the latitudes of steep vorticity  gradients  separating  parabolic retrograde jets corresponding to the well  mixed steps of the staircase \citep{Baldwin-etal-2007, Dritchel-2008}.
While vorticity staircases have been obtained in numerical simulations ~\citep{Scott-Dritchel-2012}, in many cases mixing is insufficient to produce a staircase structure. Moreover, jets are observed to form from a bifurcation at infinitesimal perturbation amplitude and in the absence of wave breaking~\citep{Farrell-Ioannou-2003-structural}.

Arguments based on equilibrium statistical mechanics  have  also been advanced to explain emergence  of jets e.g. by \cite{Miller-1990} and \cite{Robert-Sommeria-1991}. 
This theory is based on the principle that dissipationless turbulence tends to produce configurations that maximize entropy while conserving both energy and enstrophy. These maximum entropy configurations in beta plane barotropic turbulence
assume the form, when observed at large scale, of zonal jets or  vortices (cf.~\cite{Bouchet-Venaile-2012}). However, the relevance of these results to planetary 
flows that are strongly forced and dissipated and therefore out of 
equilibrium remains to be shown.

An important constraint on theories of jet maintenance is that the primary mechanism by which planetary turbulent jets are maintained is eddy momentum flux systematically directed up the mean velocity gradient; and this up-gradient momentum flux is produced by a broad spectrum of eddies, implying that the large-scale jets are maintained by spectrally nonlocal interaction between the eddy field and the large-scale zonal jets. This has been verified in observational studies on Jovian atmosphere~\citep{Ingersoll-etal-2004,Salyk-etal-2006} and in numerical simulations~\citep{Nozawa-and-Yoden-97,Huang-Robinson-98}. \citet{Wordsworth-etal-2008} studied jet formation in rotating tanks and  found strong evidence confirming that jets are maintained by non-local energy transfer. 

Laminar instability of a meridional Rossby wave or of a zonally varying 
meridional flow   can generate zonal 
fows \citep{Lorenz-1972, Gill-1974,Manfroi-Young-99,  Berloff-etal-2009a, Connaughton-etal-2010}. Equations for 
the dynamics of these jets in the weakly nonlinear  limit were obtained by
\citet{Manfroi-Young-99}. This instability,  referred to as modulational instability, produces spectrally 
nonlocal transfer to the zonal flow  from the forced meridional waves but presumes a constant 
source of these finite amplitude meridional waves. In baroclinic flows, baroclinic instability 
has been advanced as the source of these coherent waves \citep{Berloff-etal-2009a}.


Stochastic structural stability theory (SSST contracted to S3T) addresses turbulent jet dynamics as a 
two-way interaction between the mean flow and its consistent  field of turbulent 
eddies~\citep{Farrell-Ioannou-2003-structural}. Both S3T and  modulational instability  involve non-local   interactions in wavenumber space  but these theories differ in that   in S3T the mean flow is supported by its interaction 
with a broad turbulence spectrum rather than with  specific  waves. 
In fact, S3T is a non-equilibrium statistical theory that provides a closure comprising a dynamics for the evolution of
the mean flow together with its consistent field of eddies.
In S3T the dynamics of the turbulence statistics required by this closure are 
obtained from a   stochastic turbulence model (STM), which provides accurate eddy statistics for the atmosphere at large 
scale~\citep{Farrell-Ioannou-1993d, Farrell-Ioannou-1994a,Farrell-Ioannou-1995,
Zhang-Held-99}. 
 \citet{Marston-etal-2008} have shown that the S3T system is obtained by truncating the infinite hierarchy of cumulant expansions to second order and they refer to  the S3T system as the second order cumulant expansion (CE2). 
In S3T, jets initially arise as a linear instability of the interaction 
 between an infinitesimal jet perturbation and the associated eddy field and finite amplitude jets result from nonlinear equilibria continuing from these instabilities.
Analysis of this jet formation instability determines the bifurcation structure of the jet 
formation process as a function of parameters.
In addition to  jet formation bifurcations, S3T predicts  jet breakdown bifurcations 
 as well as the structure of the emergent jets, the structure of the finite amplitude 
 equilibrium jets they continue to, and the structure of the turbulence accompanying the jets.
Moreover, S3T is a dynamics so it predicts   the time dependent trajectory of the statistical 
mean turbulent state as it evolves and, remarkably, the mean turbulent state is often 
predicted by S3T to be time dependent in the sense 
that the statistical mean state of the turbulence evolves  in a 
manner predicted by the theory \citep{Farrell-Ioannou-2009-plasmas}.
The formation of zonal jets in 
planetary  turbulence was  studied as a bifurcation problem in S3T 
 by \cite{Farrell-Ioannou-2003-structural, Farrell-Ioannou-2007-structure, Farrell-Ioannou-2008-baroclinic, Farrell-Ioannou-2009-equatorial,  Farrell-Ioannou-2009-closure,Bakas-Ioannou-2011,  Srinivasan-Young-2012,Parker-Krommes-2013-generation}.
A continuous formulation of S3T developed by  \cite{Srinivasan-Young-2012} has facilitated analysis of the physical processes that give rise to the S3T instability and 
construction of  analytic expressions for the growth rates of  the S3T instability  in  homogeneous beta-plane turbulence \citep{Srinivasan-Young-2012,Bakas-Ioannou-2013-jas}.
Recently, the analogy between the dynamics of pattern formation   and zonal jet emergence  in the context of S3T was studied by~\citet{Parker-Krommes-2013-generation}.

Relating S3T  to jet dynamics in fully nonlinear turbulence is facilitated by studying the 
quasi-linear model  which is intermediate between the nonlinear model and S3T. 
The quasi-linear (QL) approximation  to the full nonlinear dynamics (NL) results when eddy-eddy interactions are not explicitly included in the dynamics but are either neglected entirely or replaced with a simple stochastic parameterization, so that no  turbulent cascade occurs in the equations for the eddies, while interaction between the eddies and the zonal mean flow is  retained fully  in the zonal mean equation.
S3T is essentially QL with the additional assumption of an infinite ensemble of eddies
replacing the single realization evolved under QL. Although the dynamics of S3T and QL 
are essentially the same, by making the approximation of an infinite ensemble of eddies,
the S3T equations  provide  an autonomous and fluctuation-free dynamics of the statistical mean turbulent state,  which transforms QL from a simulation of turbulence into a predictive theory of turbulence.

A fundamental attribute of QL/S3T is that the nonlinear eddy-eddy cascade of NL is suppressed in these systems. It follows that agreement in predictions of jet formation and equilibration  between NL and QL/S3T provides compelling evidence that cascades are not required for jet formation and theoretical support for observations showing 
that the turbulent transfers of momentum maintaining finite amplitude jets are non-local in spectral space.


%
 Previous studies demonstrated that unstable jets maintained by mean flow body forcing
 can be equilibrated using QL dynamics \citep{Schoeberl-Lindzen-84,DelSole-Farrell-1996,OGorman-Schneider-2007,Marston-etal-2008}.
%
In contrast to these studies, in this work we investigate  the spontaneous emergence and equilibration of jets from homogeneous turbulence in the absence of any coherent external forcing at the jet scale.  S3T predicts that infinitesimal perturbations with zonal jet form organize homogeneous turbulence to  produce systematic up-gradient fluxes giving rise to exponential  jet growth and eventually to the establishment  of finite amplitude equilibrium jets.  Specifically,  the S3T  equations  predict initial formation of jets by the most unstable eigenmode of the linearized S3T dynamics.
 In agreement with S3T, \cite{Srinivasan-Young-2012} found  that their NL simulations exhibit jet emergence  from a homogeneous turbulent state  with subsequent establishment of finite amplitude jets, while noting quantitative differences between
 bifurcation parameter values predicted by S3T and the parameter values for which 
 jets were observed to emerge  in NL.  \cite{Tobias-Marston-2013} also investigated the correspondence of CE2 simulations of jet formation with corresponding NL simulations and found that CE2 reproduces the jet structure, although they noted some differences in the second cumulant, and suggested a remedy by inclusion of higher cumulants.

In this paper we  use NL and its  QL counterpart together with S3T 
to examine further the dynamics  of emergence and equilibration of jets  from turbulence. 
Qualitative agreement in bifurcation behavior among these systems, which is obtained for all the spatial turbulence 
forcing distributions studied, confirms that the S3T instability mechanism is responsible for the formation and 
equilibration of jets.  Quantitative agreement is obtained for bifurcation parameters between NL and QL/S3T when account is taken of the modification of the 
turbulent spectrum that occurs in NL but not in QL/S3T.  Remarkably,  a primary component of this spectral modification can 
itself be traced to S3T instability, but of non-zonal rather than of zonal form.  We investigate 
the formation and equilibration of these non-zonal S3T instabilities and the effect these structures 
have on the equilibrium spectrum of beta-plane turbulence.  We also investigate circumstances under which 
non-zonal structures are modified and suppressed by the formation of zonal jets.

A dynamic of potential importance to climate is the possibility of multiple equilibria of the statistical mean turbulent state being supported with the same system parameters \citep{Farrell-Ioannou-2003-structural,Farrell-Ioannou-2007-structure,Parker-Krommes-2013-generation}.  
We verify existence of  multiple equilibria, predicted by S3T, in our NL simulations.
Finally, we show that weak jets result from
stochastic excitation by the turbulence of stable S3T modes, which demonstrates the physical reality of the stable S3T modes.  
Turbulent fluctuation induced excitation of these weak local jets and the weak 
but zonally extended jets that form at slight supercriticality in 
the jet instability bifurcation may explain the enigmatic latent jets of \cite{Berloff-etal-2011}.

\section{Formulation of  nonlinear and quasi-linear baro\-tropic beta-plane dynamics}

Consider a  beta-plane   with $x$ and $y$ Cartesian coordinates along  the zonal and  the meridional direction respectively. The nondivergent  zonal and meridional velocity fields are expressed  in terms of a streamfunction, $\psi$, as $u=-\partial_y \psi$ and $v=\partial_x \psi$. The absolute  vorticity  is $\z+2 \Omega+\beta y$, where 
$\z = \Delta \psi$ and $\Delta \equiv \partial^2_{xx} + \partial^2_{yy}$.
The NL dynamics of this system is governed by the barotropic vorticity equation:
\be
 \partial_t \z +u\, \partial_x \z + v\, \partial_y \z + \beta v = - r \z  -\nu_4\Delta^2\z+ \sqrt{\varepsilon} F~.
 \label{eq:q}
\ee
The flow is dissipated with linear damping at rate $r$ and hyperviscosity with coefficient $\nu_4$. 
Periodic boundary conditions are imposed in $x$ and $y$ with periodicity $2 \pi L$.  Distances have been nondimensionalized by 
$L=5000\;\textrm{km}$ and time by $T=L/U$, where $U=40\;\textrm{m\,s}^{-1}$, so that the time unit is $T=1.5\;\textrm{day}$ 
and $\beta=10$ corresponds to a midlatitude value. 
Turbulence is maintained by stochastic forcing with  spatial and temporal structure, $F$, and amplitude $\varepsilon$. 

Flow fields are decomposed  into zonal-mean components, denoted with a bar, and deviations from the zonal mean (eddies), which are indicated lowercase with primes. The zonal velocity is $U(y, t) + u'(x,y,t)$,  with $U$ the zonal mean velocity, the meridional velocity is $v'(x,y,t)$, and the eddy vorticity is $\z'(x,y,t)$. From Eq.~\eqref{eq:q},  equations for the evolution of the zonal mean flow, $U$, and the associated  eddy field, $\z'$, obtained by zonal averaging, are:
\ifthenelse{\boolean{dc}}
{\begin{subequations}
\begin{align}
\partial_t U &=  \overline{v'\z'} - r U ~,\label{eq:Ut_full}\\
\partial_t \z'&=  -U \partial_x \z'+\(\partial_{yy} U-\beta\)\partial_x \psi' -r \z'-\nu_4\Delta^2\z' +\nonumber\\
&\qquad\qquad\qquad\qquad\qquad\qquad\qquad + F_e  + \sqrt{\varepsilon} F ~,\label{eq:eddy_evol}
\end{align}
\label{eq:nl_eq}\end{subequations}
}
{\begin{subequations}
\begin{align}
\partial_t U &=  \overline{v'\z'} - r U  ~,\label{eq:Ut_full}\\
\partial_t \z'&=  -U \partial_x \z'+\(\partial_{yy} U-\beta\)\partial_x \psi' -r \z'-\nu_4\Delta^2\z' + F_e  + \sqrt{\varepsilon} F ~,\label{eq:eddy_evol}
\end{align}
\label{eq:nl_eq}\end{subequations}
}
where $F_e=  \[\bit \partial_y \( \overline{v'\z'}\) - \partial_y ( v'\z' ) \] - \partial_x (u'\z')$ is the nonlinear term representing  the eddy-eddy interactions. Equations~(\ref{eq:nl_eq}) define the nonlinear system, NL. 
Note that the stochastic forcing appears only in Eq.~\eqref{eq:eddy_evol}. 


The QL approximation of NL is obtained  by setting $F_e  = 0$, which implies neglect of the eddy-eddy interactions in Eq.~\eqref{eq:eddy_evol} while retaining the  Rey\-nolds stress forcing, $ \overline{\v'\z'}$, in  the mean zonal flow equation:
\begin{subequations}
\begin{align}
\partial_t U &=  \overline{v'\z'} - r U  ~,\label{eq:Ut_full_re}\\
\partial_t \z'&=  -U \partial_x \z'+\(\partial_{yy} U-\beta\)\partial_x \psi' -r \z'-\nu_4\Delta^2\z'   + \sqrt{\varepsilon} F ~.\label{eq:ql_full_q}
\end{align}
\label{eq:ql_eq}\end{subequations}
Both the NL system~\eqref{eq:nl_eq} and its QL counterpart~\eqref{eq:ql_eq} conserve both energy and enstrophy in the absence of forcing and dissipation. 

\section{S3T  formulation of barotropic beta-plane dynamics}

The S3T system  governs evolution of the ensemble  mean state  of the QL system (Eqs.~\eqref{eq:ql_eq}). Derivation and properties of S3T 
can be found in 
\citet{Farrell-Ioannou-2003-structural,Farrell-Ioannou-2007-structure}.
Expressed in matrix form the S3T system is:
\begin{subequations}
\begin{align}
\partial_t \Uv & = - \sum_{k=1}^{N_k} \frac{k}{2}\vecd{\[\bit \imag{(\DDel^{-1}_k \C_k)}\]} - r \Uv ~,\\
\partial_t \C_k &= \A_k(\Uv)\, \C_k + \C_k\,\A_k(\Uv)^\dag +\varepsilon \, \Q_k~,~~k=1,\dots,N_k~.\label{eq:ssst_Ck_2}\end{align}\label{eq:ssst_eq}\end{subequations}
In these equations the eddy fields and the forcing have been expanded in zonal harmonics, i.e. $\z'(x,y,t) = \real \[ \vphantom{\sum_{k=1}^{N_k}} \right.\sum_{k=1}^{N_k}$ $ \hat{\z}_k(y,t)\,e^{\i k x} \left.\vphantom{\sum_{k=1}^{N_k}}\]$, with $k=1,\dots,N_k$ the  zonal wave\-numbers. The zonal mean flow, $\Uv$, and the $k$ Fourier components of the eddy vorticity, $\skew3\hat{\zv}_k$, form column vectors with elements their corresponding values at the $N_y$ discretization points, $y_j$, $j=1,\dots,N_y$. The second order statistics of the eddy field are specified  by the $N_k$ covariances $\C_k = \<\skew3\hat{\zv}_k^{\vphantom{\dag}}\skew3\hat{\zv}^\dag_k\>$, with the angle bracket denoting an ensemble average over realizations and
$\dag$ the Hermitian transpose. The operator $\vecd{(\M)}$ returns the column vector of the diagonal elements of matrix $\M$ and  $\imag$ returns the imaginary part. The matrix $\A_k$ is: 
\be
\A_k(\Uv) = -\i k\[ \U -\(\U_{yy}-\beta\I\)\DDel^{-1}_k\]  -r\I  -\nu_4\DDel_k^2~,\label{eq:A(U)}
\ee
with $\DDel_k= \D^2-k^2 \I$, $\D^2$ the discretized $\partial_{yy}$ operator, $\I$ the identity matrix, $\DDel_k^{-1}$ the inverse of $\DDel_k$, $\U$  the  diagonal matrix with  $\vecd(\U)=\Uv$, and $\U_{yy}$ the diagonal matrix with  $\vecd(\U_{yy})=\D^2 \Uv$. 
The forcing amplitude is controlled by the parameter $\varepsilon$ and the spatial covariance of the forcing   enters Eq.~\eqref{eq:ssst_Ck_2} as $\Q_k^{\vphantom{\dag}}=\F_k^{\vphantom{\dag}}\F_k^\dag$, with $\[\bit\F_k\]_{jp} = {\rm F}_{kp}(y_j)$, where the Fourier component of the stochastic forcing, $\hat{F}_k(y,t)$, is assumed to have the form $\hat{F}_k(y,t)=\sum_{p=1}^{N_y} {\rm F}_{kp}(y)\,\xi_{kp}(t)$ (cf.~\hyperref[app:forcing]{Appendix~B}).

\section{Sto\-chastic forcing structure\label{sec:forcing_structure}}

Because the S3T instability mechanism that results in jet bifurcation from a homogeneous turbulent state differs for isotropic and non-isotropic turbulence, we consider examples of 
both isotropic and non-isotropic turbulence forcing. The jet 
forming instability in the case of homogeneous, non-isotropic 
forcing arises from the up-gradient fluxes induced by shearing of 
the turbulence by the infinitesimal perturbation jet, while the 
up-gradient fluxes for the case of homogeneous isotropic forcing arise 
from  the refraction of the eddies caused by the variation in the 
potential vorticity gradient induced by the infinitesimal perturbation jet~\citep{Bakas-Ioannou-2013-jas}.

Three stochastic forcing structures will be used in our investigation of the correspondence among S3T, QL and NL dynamics. 
The first  independently  excites a set of zonal wavenumbers.
This forcing was first used by \cite{Williams-78}  to parametrize excitation of  baro\-tropic dynamics by  baroclinic instabilities.
This forcing was also used by \cite{DelSole-01a} in his  study of  upper-level tropospheric jet dynamics and in the study of jet formation using  S3T dynamics by  \cite{Farrell-Ioannou-2003-structural,Farrell-Ioannou-2007-structure} and \cite{Bakas-Ioannou-2011}. This stochastic forcing is spatially homogeneous but not isotropic and will be denoted as NIF (non-isotropic forcing).

The second forcing, denoted IRFn, is an isotropic narrow ring forcing concentrated near a single total wavenumber. This forcing structure  has been used extensively in studies of beta-plane turbulence \citep[cf.][]{Vallis-Maltrud-93} and was also used  in the recent study of \cite{Srinivasan-Young-2012}. It was introduced by \cite{Lilly-1969}, in order to isolate  the inverse cascade from the forcing in a study of two dimensional turbulence.  The third forcing we use,
denoted IRFw, is an isotropic ring forcing in which the forcing is distributed over a wide annular region in wavenumber space around the central total wavenumber.

\begin{figure}[h!]
\centering
\includegraphics[width=19pc,trim=15mm 28mm 20mm 5mm,clip]{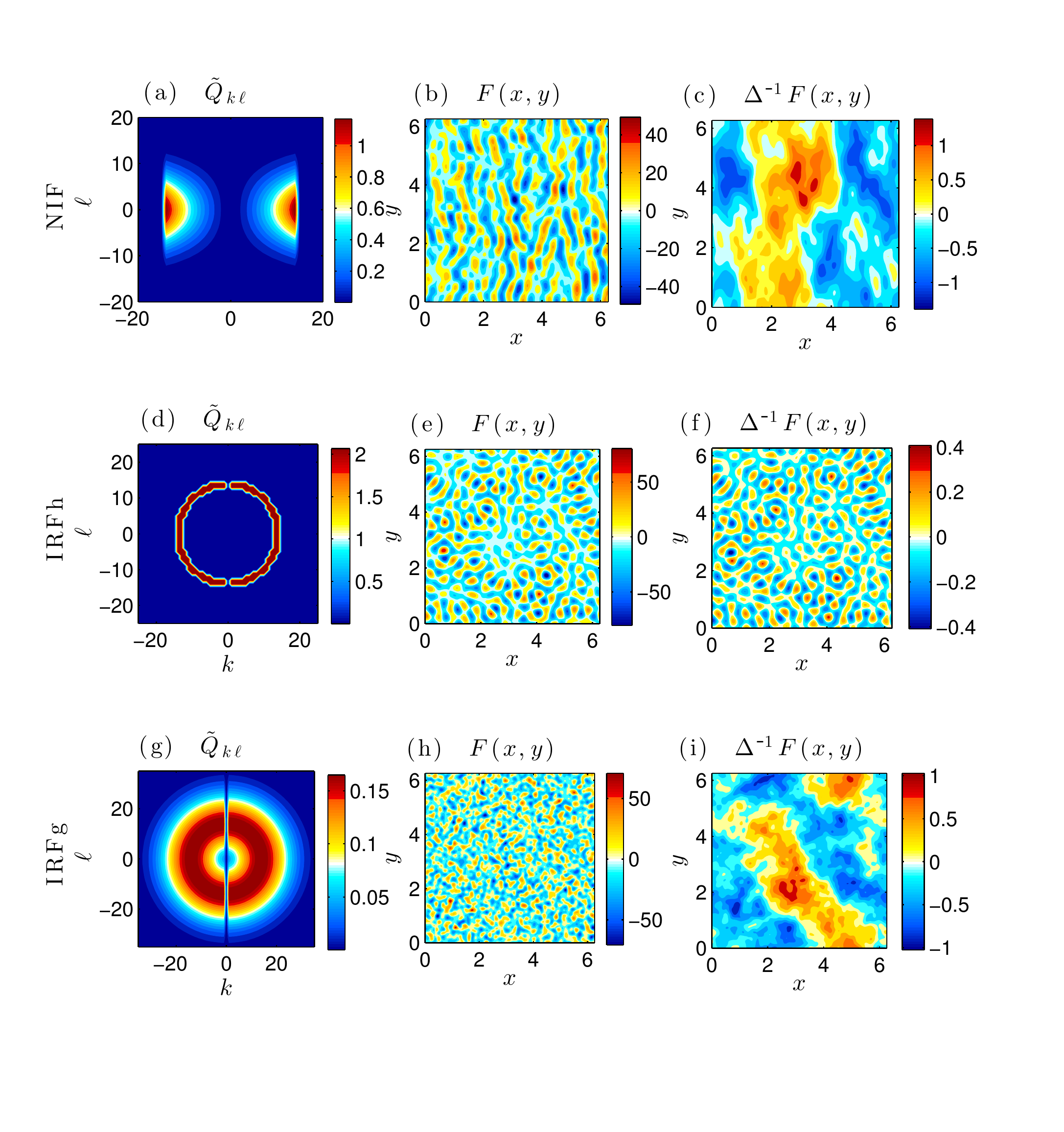}
\vspace{-2mm}\caption{Contour plots of the spatial Fourier coefficients of the forcing vorticity covariances, $\tilde{Q}_{k\l}$ (cf.~Eq.~\eqref{eq:Q_ab_to_Q_kl_posneg}),
used in this study and example realizations of the forcing.
Panel (a):  $\tilde{Q}_{k\l}$ for NIF with zonal wavenumbers $k=1,\dots,14$ and $s=0.2/\sqrt{2}$. 
Panel (d): $\tilde{Q}_{k\l}$  for IRFn  at  $K_f=14$ and $\d k_f=1$.
Panel (g) $\tilde{Q}_{k\l}$  for IRFw  at  $K_f=14$ and $\d k_f=8/\sqrt{2}$.
In (b), (e) and (h)  are shown realizations of these forcings in the vorticity field, 
and in  (c), (f) and (i)  are shown realizations in the streamfunction field.
 }
  \label{fig:stripe_ring_Qkl}
\end{figure}

Specification of these stochastic forcing structures are given in \hyperref[app:forcing]{Appendix~B}. Plots of the corresponding power spectra together with instantaneous realizations both in vorticity and streamfunction for the three types of forcing structures are shown in Fig.~\ref{fig:stripe_ring_Qkl}.
The IRFn ring forcing is peculiar in that it primarily excites vortices of scale $1/K_f$  that are evident in both the vorticity and streamfunction fields,  while IRFw produces a streamfunction field dominated by  large scale structure similar to the fields excited by the other broadband forcings.

\begin{figure}[h!]
\centering
\includegraphics[width=16pc]{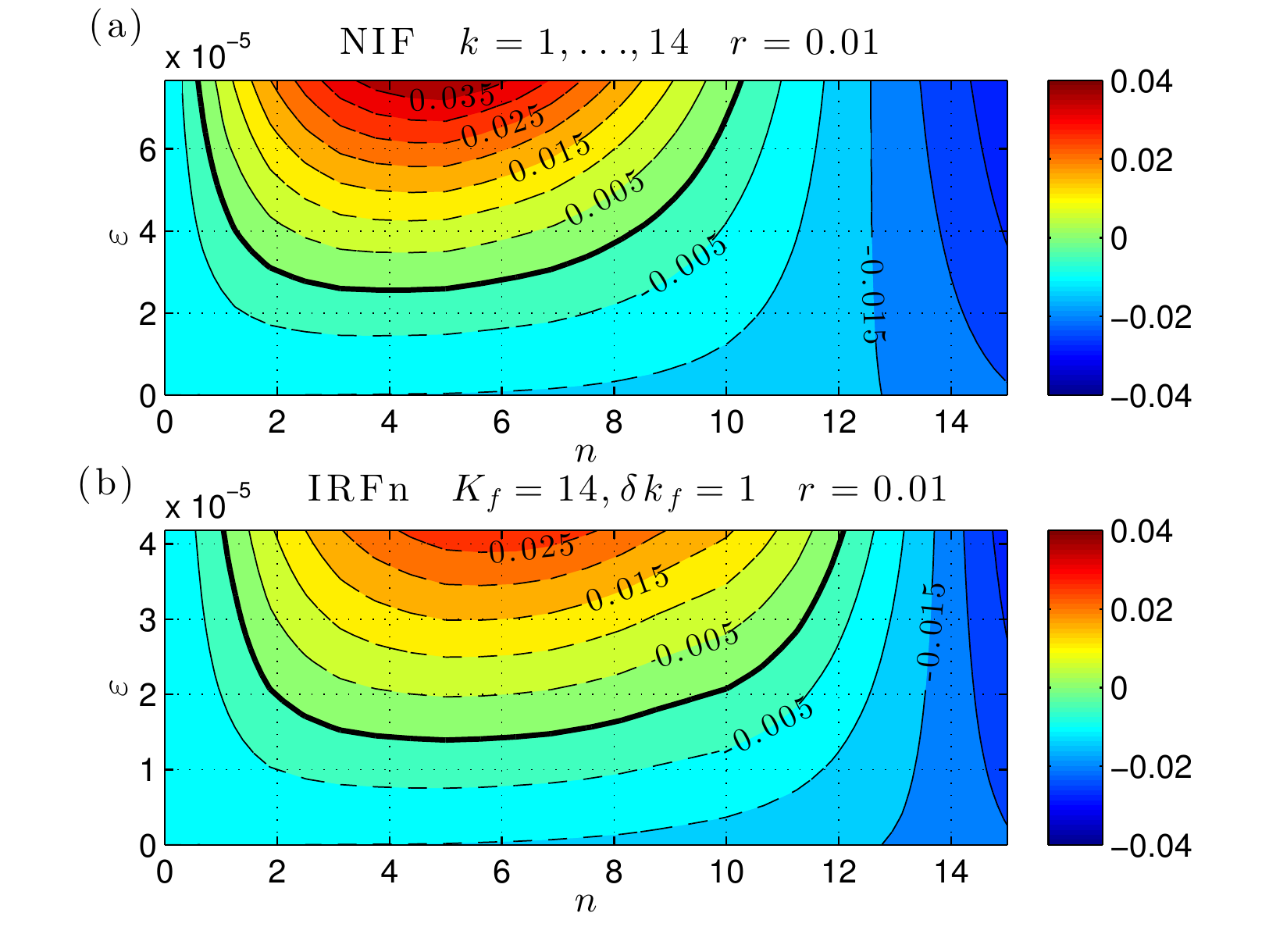}
\vspace{-2mm}\caption{Growth rate, $\sigma_r$, of the S3T eigenfunction with jet structure $\d U_n = \sin{(ny)}$  as a function of meridional wavenumber, $n$, and energy input rate, $\varepsilon$, for NIF (panel (a)) and for IRFn (panel (b)).  
 The stability boundary ($\s_r=0$) is marked with thick solid line. For NIF the
 instability occurs at $n=4$ and 
 the energy input rate required for instability is $\varepsilon_c=2.56\times10^{-5}$.
 For IRFn the instability occurs at $n=5$ and 
 the energy input rate required for instability is $\varepsilon_c=1.40\times10^{-5}$. 
 Parameters: $r=0.01$ and $\beta=10$.}
   \label{fig:S3Tgr_NIF_IRFh_r0p01}
\end{figure}

\section{Stability in S3T of the homogeneous equilibrium state}

The S3T system with homogeneous stochastic forcing and $\nu_4=0$ admits the homogeneous equilibrium solution:\be
\Uv^E=0~~~,~~~\C^E= \sum_{k=1}^{N_k} \C_k^E ~~~{\rm with }~~~\C_k^E = \frac{\varepsilon}{2r} \Q_k ~,
\label{eq:UeCe}
\ee
which has no jets and  eddy covariance proportional to the  covariance of the forcing. 
The linear stability of perturbations  $(\d\Uv ,\d \C_1, \dots ,\d \C_{N_k})$ to $(\Uv^E,\C^E)$ is determined from  the linearized equations: 
\begin{subequations}
\begin{align}
\partial_t\, \d \Uv &=-  \sum_{k=1}^{N_k} \frac{k}{2}\vecd{\[\bit \imag{(\DDel^{-1}_k \d\C_k)}\]} - r\, \d \Uv ~,\label{eq:ssst_eq_pertU_dt}\\
\partial_t\, \d \C_k &= \A^E_k\,\d\C_k + \d\C_k \(\A^E_k\)^\dag+ \d\A_k \,\C^E_k + \C^E_k\(\d\A_k\)^\dag ~,\label{eq:ssst_eq_pertCk_dt}
\end{align}\label{eq:ssst_eq_pert_dt}\end{subequations}
with $\A_k^E \equiv \A_k(\Uv^E)$ and $\d\A_k =-\i k\[ \d\U -(\d\U)_{yy}\,\DDel^{-1}_k\]$.  
The temporal eigenvalues, $\sigma$, of these linear equations,
which determine the stability of the equilbrium $(\Uv^E,\C^E)$,   satisfy Eqs.~\eqref{eq:hats} of \hyperref[app:stability]{Appendix~C}. 
 We obtain the stability of the homogeneous equilibrium  under NIF and IRFn  with $r=0.01$ 
as a function of the  parameter $\varepsilon$, which corresponds to the non-dimensional rate of energy injection into the system~(cf.~\hyperref[app:forcing]{Appendix~B}). 
The growth rates, $\s_r=\real{(\sigma)}$, as a function of the meridional wavenumber of the mean zonal flow perturbation, $n$,  and of the energy input rate, $\varepsilon$,  
obtained  using the method discussed in 
\hyperref[app:stability]{Appendix~C}, are shown in Fig.~\ref{fig:S3Tgr_NIF_IRFh_r0p01}. 
For both types of forcing, instability occurs for $\varepsilon>\varepsilon_c$ and over a band of 
mean zonal flow wavenumbers, $n$. In all cases the $\sigma$ with greatest real part has zero imaginary part implying non-translating jets have the largest growth rate. In the next section predictions of S3T stability analysis for the bifurcation structure associated with jet formation will be compared  with the corresponding  QL and NL simulations.
While QL and NL simulations reveal an apparent bifurcation, they can not provide theoretical 
predictions of this bifurcation. We wish to examine the circumstances under which the 
underlying bifurcation structure predicted theoretically by the S3T stability analysis is reflected in the QL and NL realizations.


\section{Bifurcations  predicted by S3T and their reflection in  QL  and NL simulations}

\begin{figure}[h!]
\centering
\includegraphics[width=19pc,trim=0mm 16mm 0mm 5mm,clip]{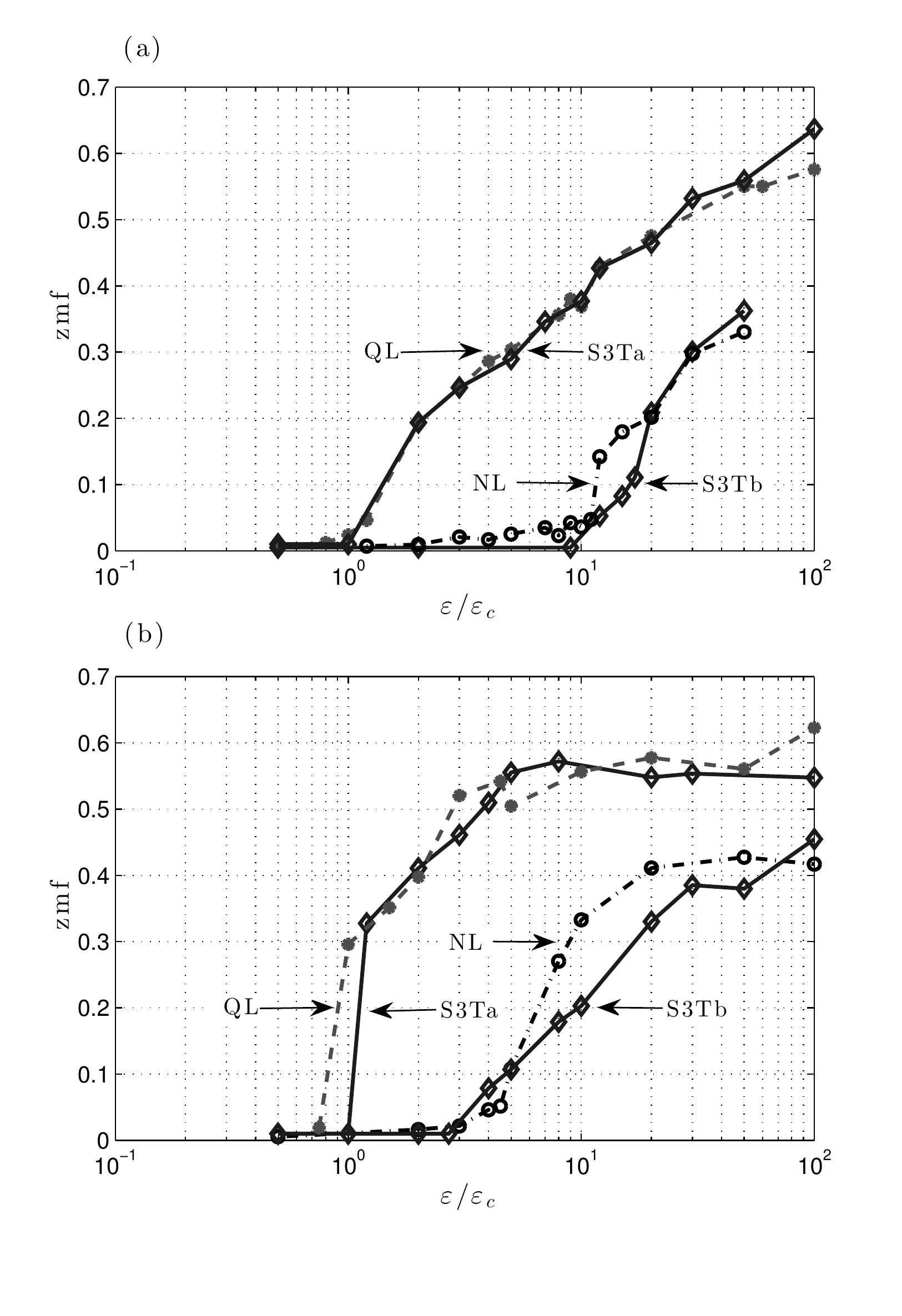}
 \vspace*{-1mm} 
 \caption{Bifurcation structure comparison for jet formation in S3T, QL, and NL. 
 Shown is the zmf index of  jet equilibria  for  NIF (panel (a)) and for IRFn (panel (b)) as a function of the forcing amplitude $\varepsilon/\varepsilon_c$ for the NL simulation  (dash-dot and circles), the QL simulation (dashed and dots) and  the corresponding  S3Ta simulation (solid). The bifurcation diagram and the structure of the jet agree in the QL and S3Ta simulation, but  the bifurcation in the NL simulations occurs at  $\varepsilon_c^{(\textrm{NL})}\approx 11\varepsilon_c$ for NIF and at at $\varepsilon_c^{(\textrm{NL})}\approx 4 \varepsilon_c$ for IRFn.  
 Agreement between NL and S3T predictions is obtained if the S3T is forced with the spectrum that reflects the modification of the  equilibrium NIF or IRFn spectrum respectively by eddy-eddy interactions (the results of this S3T simulation is indicated as S3Tb, see discussion at section~\ref{sec:8}). (For IRFn this spectrum is shown in Fig.~\ref{fig:NLspect_stability_r0p01}\hyperref[fig:NLspect_stability_r0p01]{c}.) This figure shows that the structural stability of jets in NL simulations is captured by the S3T if account is taken of the nonlinear modification of the spectrum. Parameters: $\beta=10$, $r=0.01$. } 
  \label{fig:bifurc_NIF_IRFh_r0p01}
\end{figure}

\begin{figure}[h!]
\centering
\includegraphics[width=19pc]{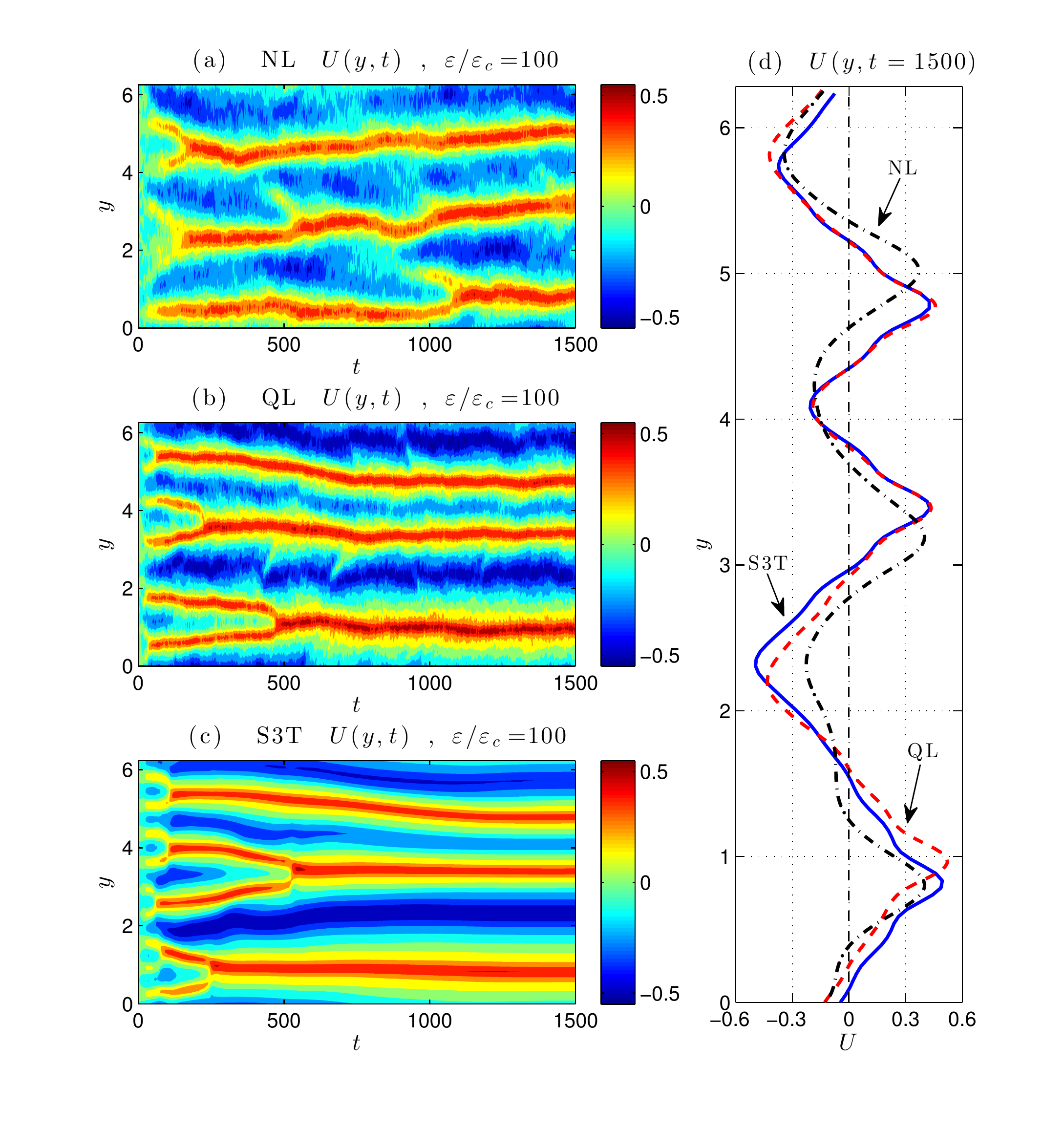}
\caption{Hovm\"oller diagrams of  jet  emergence  in  NL, QL and S3T simulations with IRFn forcing  at energy input rate $\varepsilon= 100\varepsilon_c$. Shown is $U(y,t)$ for the NL (panel (a)), QL (panel (b)) and S3T (panel (c)) simulations. Also shown are the equilibrium jets (panel (d)) in the NL (dash-dot), QL (dashed), and S3T (solid) simulations. There is very good agreement between the jet structure in the NL, QL and  S3T simulations, despite the difference in the zmf index among them (cf.~Fig.~\ref{fig:bifurc_NIF_IRFh_r0p01}\hyperref[fig:bifurc_NIF_IRFh_r0p01]{b}). Moreover, in all three simulations similar jet mergers are observed, leading eventually to final equilibrium  jets with smaller meridional wavenumber than that of the initial instability. Parameters are $\beta=10$, $r=0.01$.}
\label{fig:hov_IRFh_e100ec_r0p01}
\end{figure}

We examine the counterpart  in NL and QL simulations of the S3T structural instability by comparing the evolution of the domain averaged energy of the zonal flow, $E_m(t) = (L_xL_y)^{-1} \iint \df x\,\df y\,U^2/2$. The amplitude of the  zonal flow is measured,
as in \citet{Srinivasan-Young-2012},  with the zonal mean flow index (zmf) defined as $\textrm{zmf}=E_m/(E_m+E_p)$, where $E_m$ is the time averaged energy of the zonal mean flow, $E_m(t)$ and $E_p$ is the time average of the domain averaged kinetic energy of the eddies, $E_p(t) = (L_xL_y)^{-1}  \iint \df x\,\df y\,\(u'\skew2^2 + v'\skew2^2\)/2$.

Zonal flow (zmf) indices  are shown as a function of $\varepsilon$  in  Fig.~\ref{fig:bifurc_NIF_IRFh_r0p01}\hyperref[fig:bifurc_NIF_IRFh_r0p01]{a} for NIF forcing and in Fig.~\ref{fig:bifurc_NIF_IRFh_r0p01}\hyperref[fig:bifurc_NIF_IRFh_r0p01]{b} for IRFn forcing, both for the case with $r=0.01$ presented in Fig.~\ref{fig:S3Tgr_NIF_IRFh_r0p01}. The fundamental qualitative prediction of S3T that jets form as a bifurcation in the strength of the turbulence forcing is verified in these plots. Agreement in the  critical
value, $\epsilon_c$, for jet emergence is also obtained between S3T and QL while this parameter value is substantially larger in NL. For example, jets emerge in the NL simulations at $\varepsilon_c^{(\textrm{NL})}\approx11\varepsilon_c$ under NIF forcing and at $\varepsilon_c^{(\textrm{NL})}\approx 4\varepsilon_c$ under IRFn forcing. Similar behavior was noted by \cite{Srinivasan-Young-2012}. The  reason for this difference  will be  explained in section~\ref{sec:8}.

S3T dynamics not only predicts the emergence of zonal jets as a bifurcation in turbulence forcing, but also predicts the structure of the finite amplitude jets that result from equilibration of the initial jet formation instability. These finite amplitude jets correspond  to fixed points of the S3T dynamics. An example  for IRFn strongly forced with $\varepsilon=100 \varepsilon_c$ and with damping $r=0.01$ is shown in Fig.~\ref{fig:hov_IRFh_e100ec_r0p01}. This example demonstrates the essential similarity among the jets in NL, QL and S3T simulations.

\begin{figure}[h!]
\centering
\includegraphics[width=19pc]{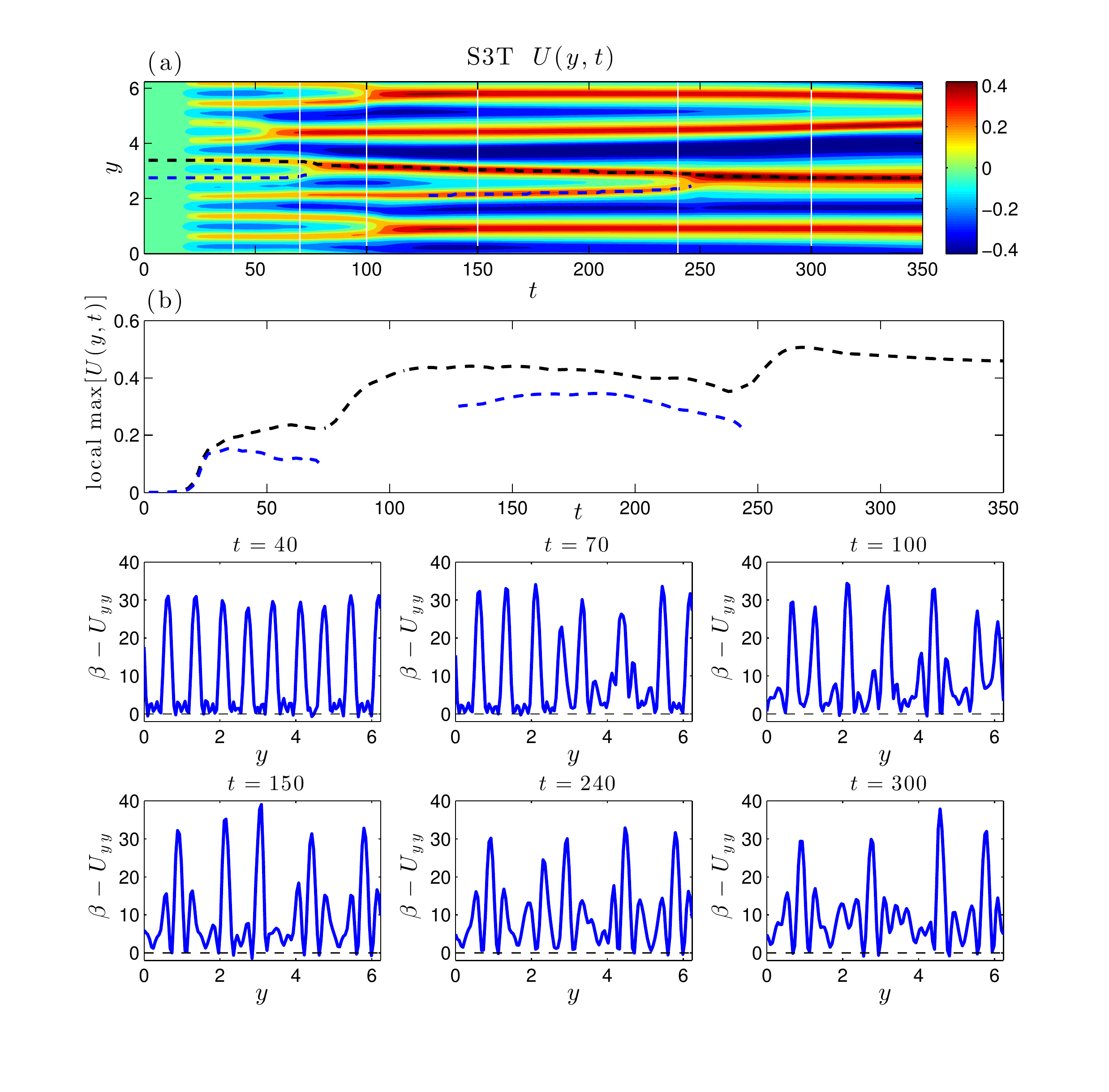}
\vspace{-2mm}
\caption{(a) Hovm\"oller diagram showing details of the jet mergers for  $t\le 350$ in the S3T simulation  in  Fig.~\ref{fig:hov_IRFh_e100ec_r0p01}. 
In (b) is shown the amplitude of the jet maxima that appear in (a). Note that only the prograde jets merge. The bottom panels show the mean potential vorticity gradient $\beta-U_{yy}$ as a function of $y$ at the times indicated by  vertical lines in (a). These graphs show that the structure of the jets is configured at each instant to satisfy  the  Rayleigh-Kuo stability criterion  and that jet mergers are the mechanism in S3T for avoiding  inflectional instability. Decrease in the amplitude of the jets prior to  merger indicates increased downgradient vorticity fluxes as the flow approaches hydrodynamic neutrality.}
\label{fig:mergers}
\end{figure}

\begin{figure}[h!]
\centering
\includegraphics[width=19pc,trim=0mm 20mm 0mm 3mm,clip]{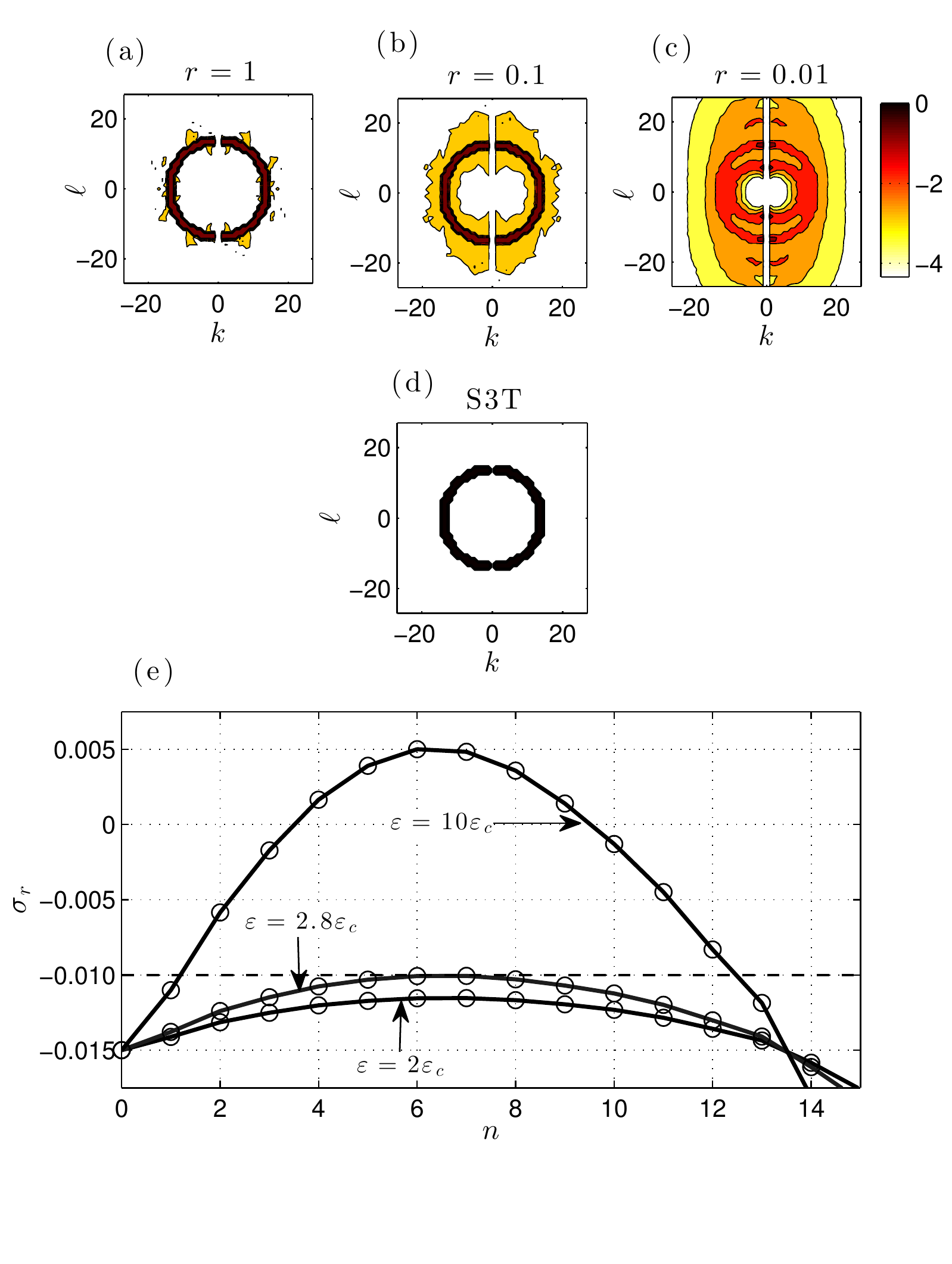}
  \caption{Panels (a)-(d): Equilibrium enstrophy spectrum, $\log{\( \<|\tilde{\z}_{k\l}|^2\>\)}$,   of NL simulations, in which eddy-eddy interactions are included  and 
the  $k=0$ component is excluded,  for various damping rates, $r$. The example is for IRFn forcing at $\varepsilon=2 \varepsilon_c$. Shown are spectra for: (a)  $r=1$, (b) $r=0.1$ and (c) $r=0.01$. The critical $\varepsilon_c$ is a function of $r$ and is obtained  from  S3T for each value of $r$. All spectra have been normalized. The equilibrium spectrum of the S3T (identical to QL) is shown in panel (d). This figure shows that for  strong damping the spectrum in NL simulations is close to the S3T spectrum while for weak damping the equilibrium spectrum in NL differs substantially from that in S3T. In all cases $\beta=10$. Panel (e): S3T  growth rates, $\s_r$, as a function of the meridional wavenumber, $n$, for the nonlinearly modified spectrum shown in panel (c) ($r=0.01$). Shown are cases for  $\varepsilon=2\varepsilon_c$, $\varepsilon=2.8\varepsilon_c$ and $\varepsilon=10\varepsilon_c$. It can be seen that S3T stability analysis forced by this spectrum predicts that  jets  should emerge at $\varepsilon = 2.8 \varepsilon_c$ with $n=6$. S3T predictions are verified in NL as shown in the bifurcation diagram in Fig.~\ref{fig:bifurc_NIF_IRFh_r0p01}\hyperref[fig:bifurc_NIF_IRFh_r0p01]{b} (denoted as  S3Tb).}  
  \label{fig:NLspect_stability_r0p01} 
\end{figure}

\begin{figure}[h!]
\centering
\includegraphics[width=19pc,trim=0mm 7mm 0mm 1mm,clip]{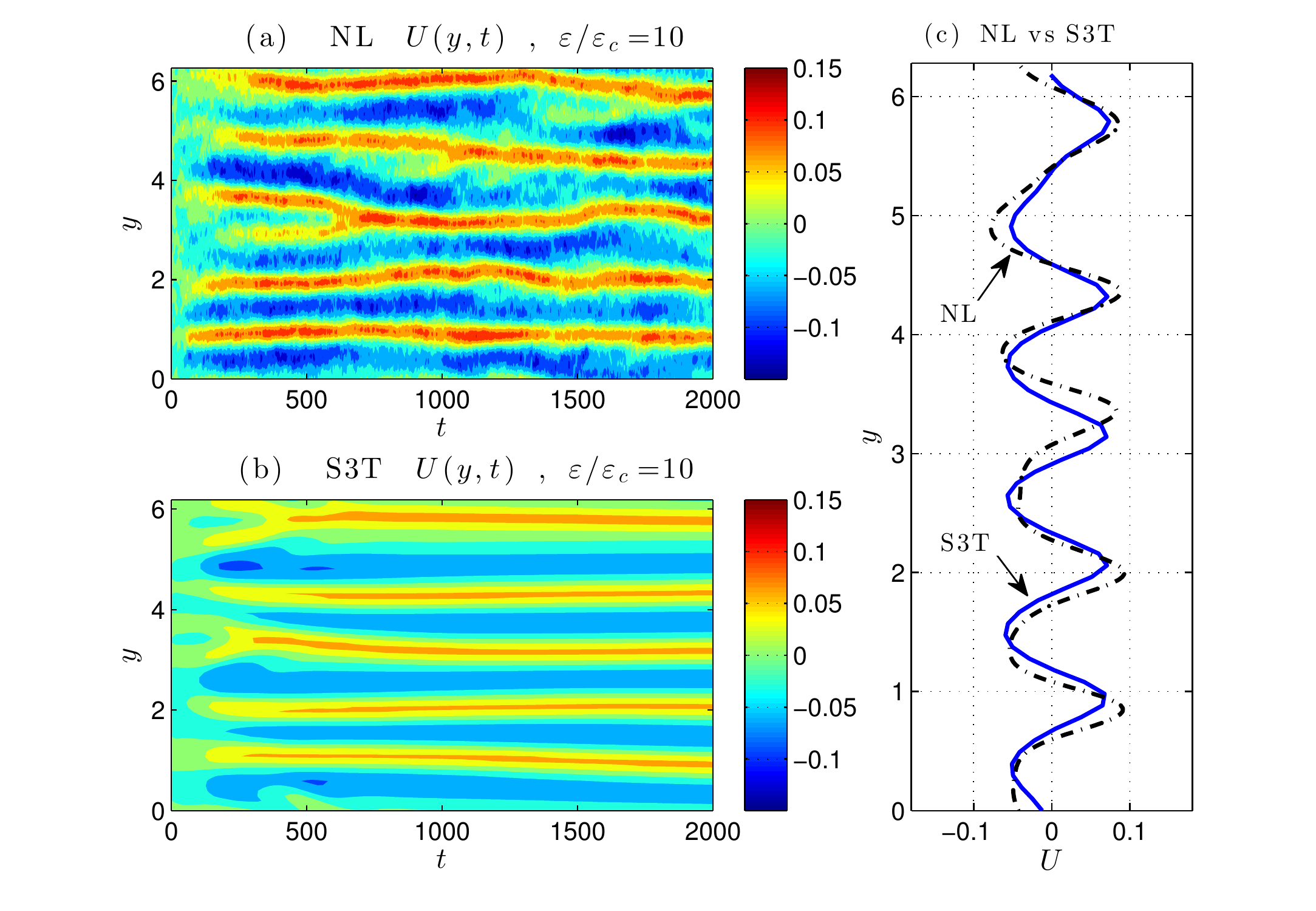}
  \caption{Hovm\"oller diagrams of $U(y,t)$ comparing  jet emergence and equilibration in an NL 
  simulation under IRFn forcing (panel (a))  with  an S3T simulation  under S3Tb forcing  (panel (b)). The corresponding time mean jets are shown in panel (c). This figure shows that  the S3Tb modification of the forcing spectrum suffices to obtain  agreement with NL. Parameters  are  $\varepsilon=10\varepsilon_c$, $\beta=10$, $r=0.01$.}  
  \label{fig:hov_IRFh_e10ec_S3Tb} 
\end{figure}

\begin{figure}[h!]
\centering\includegraphics[width=19pc,trim=0mm 13mm 0mm 1mm,clip]{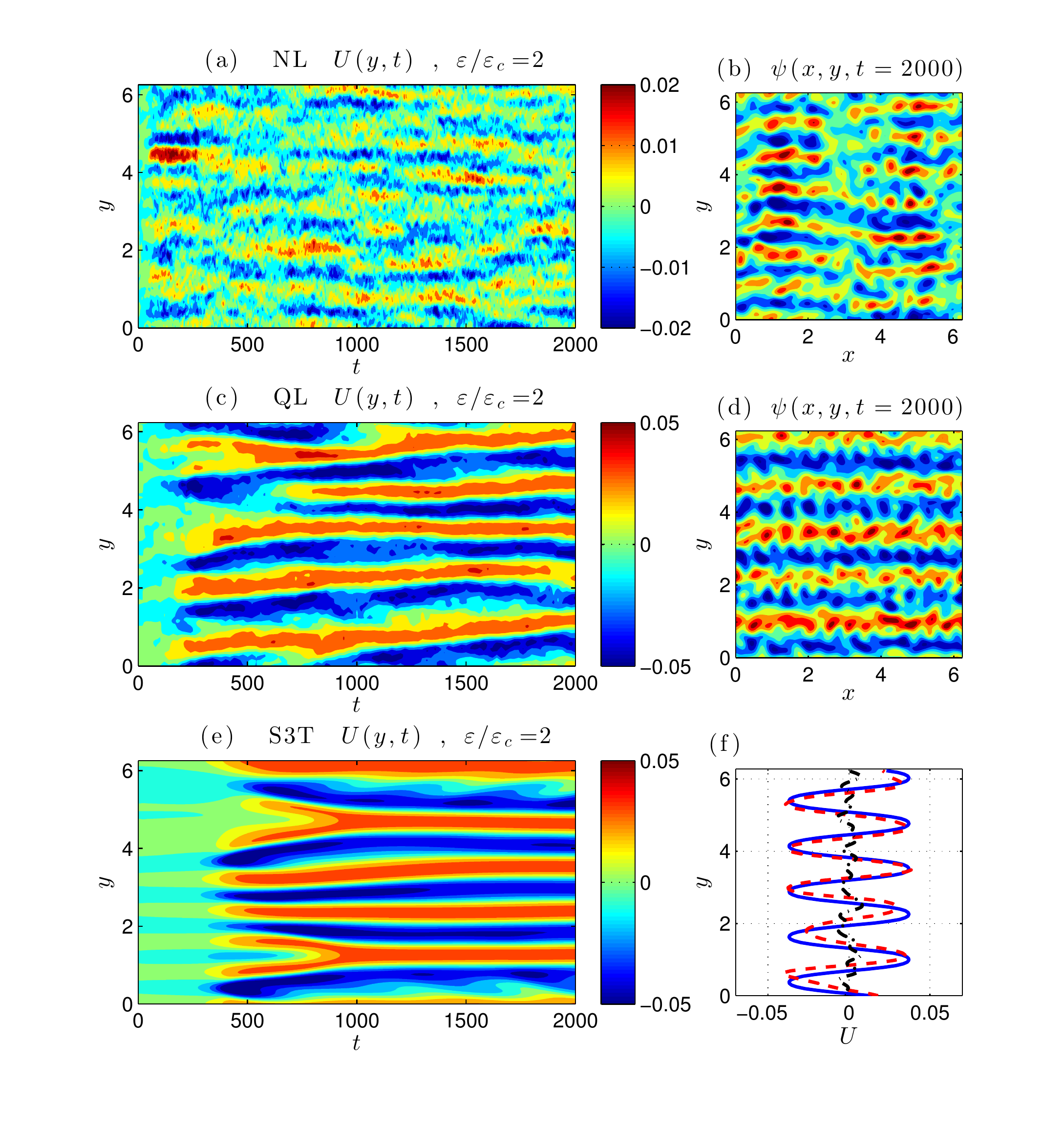}
\caption{Hovm\"oller diagrams of  jet  emergence  in  NL, QL and S3T  simulations with IRFn forcing  at  $\varepsilon= 2 \varepsilon_c$.  Shown is $U(y,t)$ for the NL (panel (a)), QL (panel (c)) and S3T (panel (e)) simulations and characteristic snapshots of streamfunction fields at $t=2000$ for the NL and QL simulations (panels (b) and (d)). Notice that in the $U(y,t)$ diagram for NL the color axis is scaled differently. Also shown are the equilibrium jets in the NL (dash-dot), QL (dashed), and S3T (solid) simulation (panel (f)).  
 At $\varepsilon=2\varepsilon_c$ in the NL simulation no jets emerge
 but accumulation of energy in $(1,7)$ non-zonal structures with zonal wavenumber $k=1$
 and meridional wavenumber $\l=7$ is discernible. Parameters are $\beta=10$, $r=0.01$.}
 \label{fig:hov_IRFh_e2ec_r0p01}
\end{figure}

Under strong turbulence forcing the initial S3T jet formation  instability typically reaches final equilibrium  as a finite amplitude jet at a wavenumber smaller than that of the initial instability. 
An example is the case of  IRFn at $\varepsilon=100\varepsilon_c$ shown in Fig.~\ref{fig:hov_IRFh_e100ec_r0p01}.
In this example,  the jets  emerge in S3T initially with zonal wavenumber $n=10$, in agreement with the prediction of the S3T instability of the homogeneous equilibrium, but eventually equilibrate at wavenumber $n=3$ following a series of jet mergers, as seen in the Hovm\"oller diagram. Similar dynamics are evident in the NL and QL simulations.
This behavior can be rationalized by noting that if the wavenumber of the jet remains fixed then as jet amplitude continues to increase under strong turbulence forcing violation of the Rayleigh-Kuo stability criterion would necessarily occur. By transitioning to a lower wavenumber the flow is able to forestall this occurrence of inflectional instability. However, detailed analysis of the S3T stability of the finite amplitude equilibria  near the point of jet merger reveals that these mergers coincide with the inception of a structural instability associated with 
eddy/mean flow interaction, which precedes the occurrence  of hydrodynamic instability of the jet~\citep{Farrell-Ioannou-2003-structural,Farrell-Ioannou-2007-structure}\footnote{
Jet mergers occur  in the  Ginzburg-Landau equations  that govern the dynamics of  the S3T instability of the homogeneous equilibrium state  for parameter values for which the system is close to marginal stability \citep{Parker-Krommes-2013-generation}. 
However,  these mergers in the Ginzburg-Landau equations are  associated with  equilibration  of the Eckhaus instability, rather than equilibration of the inflectional instability associated with violation of the Rayleigh-Kuo criterion as is the case for mergers of finite amplitude jets. Characteristic of this difference is that in the case of the 
Ginzburg-Landau equations both the prograde and retrograde jets merge, while in the case of the finite amplitude jets only the prograde jets merge. The same phenomenology as in the Ginzburg-Landau equations occurs in the case of the
Cahn-Hilliard equations that govern the dynamics of marginally stable jets in the modulational instability 
study of \cite{Manfroi-Young-99}. }.

\section{Influence of the turbulence spectrum on the S3T jet formation instability\label{sec:8}}

Both QL and S3T dynamics  exclude interactions among eddies  and include only the  non-local interactions between jets, with $k=0$, and eddies, with $k\ne0$. Therefore, there is no  enstrophy or energy cascade in wavenumber space in either QL or S3T dynamics and the homogeneous S3T equilibrium state \eqref{eq:UeCe} has spectrum,  ${\varepsilon}\tilde{Q}_{k\l} / (2 r)$,which is determined  by the spectrum of the forcing ($\tilde{Q}_{k\l} $ is the spectral power of the forcing covariance, cf. \hyperref[app:forcing]{Appendix~B}).
However, this is not true in NL  which includes eddy-eddy interactions  producing enstrophy/energy cascades. 
For example, in NL an isotropic ring forcing  is spread  as time progresses,  becoming concentrated at lower wavenumbers and forming the characteristic dumbbell shape seen in  beta-plane turbulence simulations (cf.~\cite{Vallis-Maltrud-93}) and  consequently   the homogeneous turbulent state is no longer characterized by the spectrum of the forcing. We can take account of this modification of the spectrum by performing S3T stability on the homogeneous state under the equivalent forcing covariance,
\be
\tilde{Q}_{k\l}^\textrm{NL} =  \frac{2 r}{\varepsilon}\,\<\bit |\tilde{\z}_{k\l}|^2\>~.\label{eq:nlspectrum}
\ee
which maintains the observed NL spectrum,  $\< \bit  |\tilde{\z}_{k\l}|^2\> $, in the S3T dynamics. The NL modified eddy vorticity spectrum, $\< \bit  |\tilde{\z}_{k\l}|^2\> $, is obtained from an ensemble of NL simulations.
Plots of  $\< \bit  |\tilde{\z}_{k\l}|^2\>$, under IRFn forcing are shown in Figs.~\ref{fig:NLspect_stability_r0p01}\hyperref[fig:NLspect_stability_r0p01]{a-c} for 
various energy input rates, $\varepsilon$, and damping rates, $r$. 
The departure of the NL spectra from the spectra of the QL and S3T equilibria is evident and this 
departure depends on the amplitude of the forcing, $\varepsilon$, and the damping, $r$.

We now demonstrate that while the fundamental qualitative prediction of S3T that jets form as a bifurcation in turbulence forcing and in the absence of turbulent cascades is verified in both QL and NL, a necessary condition for obtaining quantitative agreement between NL and both S3T and QL dynamics is that  the equilibrium spectrum used in the S3T and QL dynamics be close to the equilibrium spectrum obtained in NL so that the stability analysis is performed on similar states. 
In the case with IRFn and  $r=0.01$, formation of persistent finite amplitude zonal jets occurs in the NL simulations  at $\varepsilon=2.8 \varepsilon_c$ (cf.~Fig.~\ref{fig:bifurc_NIF_IRFh_r0p01}\hyperref[fig:bifurc_NIF_IRFh_r0p01]{b}).  
In agreement, S3T stability analysis  on the NL modified equilibrium IRFn spectrum (denoted S3Tb and shown in Fig.~\ref{fig:NLspect_stability_r0p01}\hyperref[fig:NLspect_stability_r0p01]{c})
predicts instability for  $\varepsilon \ge 2.8 \varepsilon_c$  (cf.~Fig.~\ref{fig:NLspect_stability_r0p01}\hyperref[fig:NLspect_stability_r0p01]{e}).
Moreover, S3T stability analysis with the S3Tb spectrum predicts  jet 
formation at $n=6$ and in agreement with this prediction jets  emerge in NL with $n=6$. 
Hovm\"oller diagrams  demonstrating  similar jet evolution in NL under 
IRFn forcing  and in S3T under S3Tb  forcing are shown in Fig.~\ref{fig:hov_IRFh_e10ec_S3Tb}. We also note that
agreement between NL and S3T in predictions of  jet amplitude at large supercriticality is 
also obtained by using the S3Tb spectrum (cf. Figs.~\ref{fig:bifurc_NIF_IRFh_r0p01}\hyperref[fig:bifurc_NIF_IRFh_r0p01]{a}~and~\ref{fig:bifurc_NIF_IRFh_r0p01}\hyperref[fig:bifurc_NIF_IRFh_r0p01]{b})\footnote{The spectral peaks near the $\l$ axis do not directly influence  the stability of the NL modified spectrum, which is determined by the distorted and broadened ring spectrum. However, while the spectral peaks do not influence the stability directly, they do
influence it indirectly by distorting the incoherent spectrum.}.

\begin{figure}[h!]
\centering\includegraphics[width=19pc]{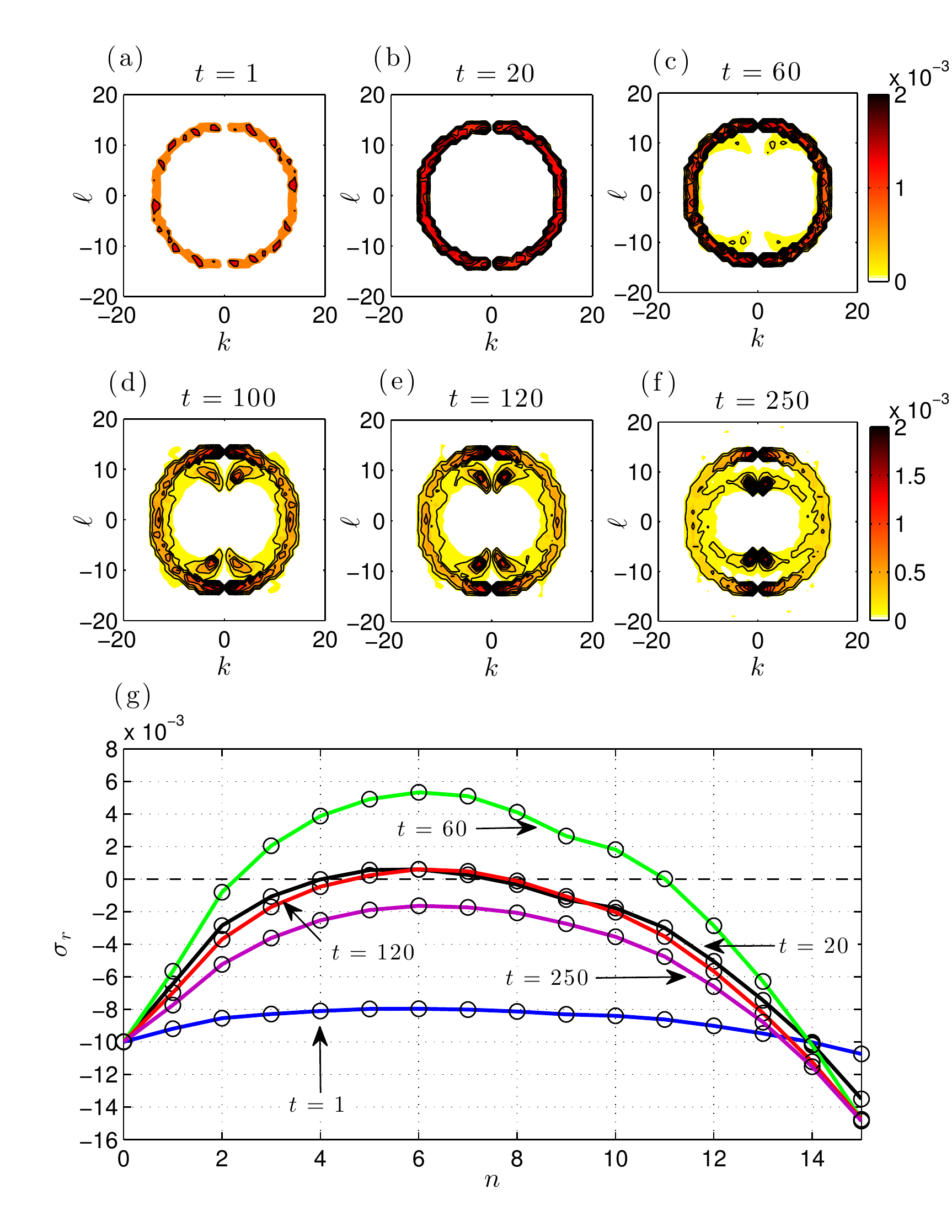}
\caption{Panels (a)-(f): Evolution of the ensemble average enstrophy spectrum, $\<|\tilde{\z}_{k\l}|^2\>$, for NL  with IRFn forcing at $\varepsilon=2\varepsilon_c$. Panel (g): Growth rates, $\s_r$,  as a function of jet meridional wavenumber, $n$,  predicted by S3T stability analysis performed on the instantaneous  spectrum at the times indicated  in panels (a)-(f). The evolving spectrum renders the NL simulation  S3T unstable at $t\approx20$ and stabilizes it again at $t\approx120$. Parameters are  $\beta=10$, $r=0.01$.} 
  \label{fig:NL_spectr_stability}
\end{figure}

This influence of the eddy spectrum on jet dynamics is revealed in the case of IRFn at
energy input rate $\varepsilon=2\varepsilon_c$, shown in Fig.~\ref{fig:hov_IRFh_e2ec_r0p01}. Although at this energy input rate S3T under IRFn  is structurally unstable, no jets emerge in NL. We have shown that agreement in  bifurcation structure is obtained between NL and S3T when S3T analysis 
is performed with  the S3Tb spectrum. We now examine the development of the NL spectrum  towards S3Tb and demonstrate the close control  exerted by this evolving spectrum on S3T stability.  The evolving spectrum, shown in Fig.~\ref{fig:NL_spectr_stability}\hyperref[fig:NL_spectr_stability]{a-f}, is obtained using an ensemble of NL simulations, each starting from a state of rest and evolving under a different forcing realization.  A sequence of S3T stability analyses performed on this evolving ensemble spectrum is show in Fig.~\ref{fig:NL_spectr_stability}\hyperref[fig:NL_spectr_stability]{g}. The weak NL ensemble spectrum at $t=1$ does not support instability, but by $t=20$ the ensemble spectrum, having assumed  the isotropic ring structure of the forcing, becomes  S3T unstable.
 This structural instability results in the formation of an incipient $n=6$ jet structure which is evident by $t=50$ in the NL simulation  shown  in Fig.~\ref{fig:hov_IRFh_e2ec_r0p01}.
As the spectrum further evolves, the S3T growth rates decrease and no jet structure is unstable for $t > 120$, and decay rates continue to increase 
until $t=250$   (cf.~Fig.~\ref{fig:NL_spectr_stability}\hyperref[fig:NL_spectr_stability]{g}). 
This example demonstrates the tight control on S3T stability exerted by the spectrum. 
Furthermore, it shows the close association between S3T instability and the  emergence of jet structure in NL.

%

\ifthenelse{\boolean{dc}}
{\section{Influence  of $\,\!$ non-zonal structures predicted by S3T on the turbulence spectrum and on jet dynamics \label{sec:nonzonal}}
}
{
\section{The influence  of $\,\!$ non-zonal  structures  predicted by S3T on the turbulence spectrum and on jet dynamics\label{sec:nonzonal}}
}

\begin{figure}[h!]
\centering\includegraphics[width=19pc]{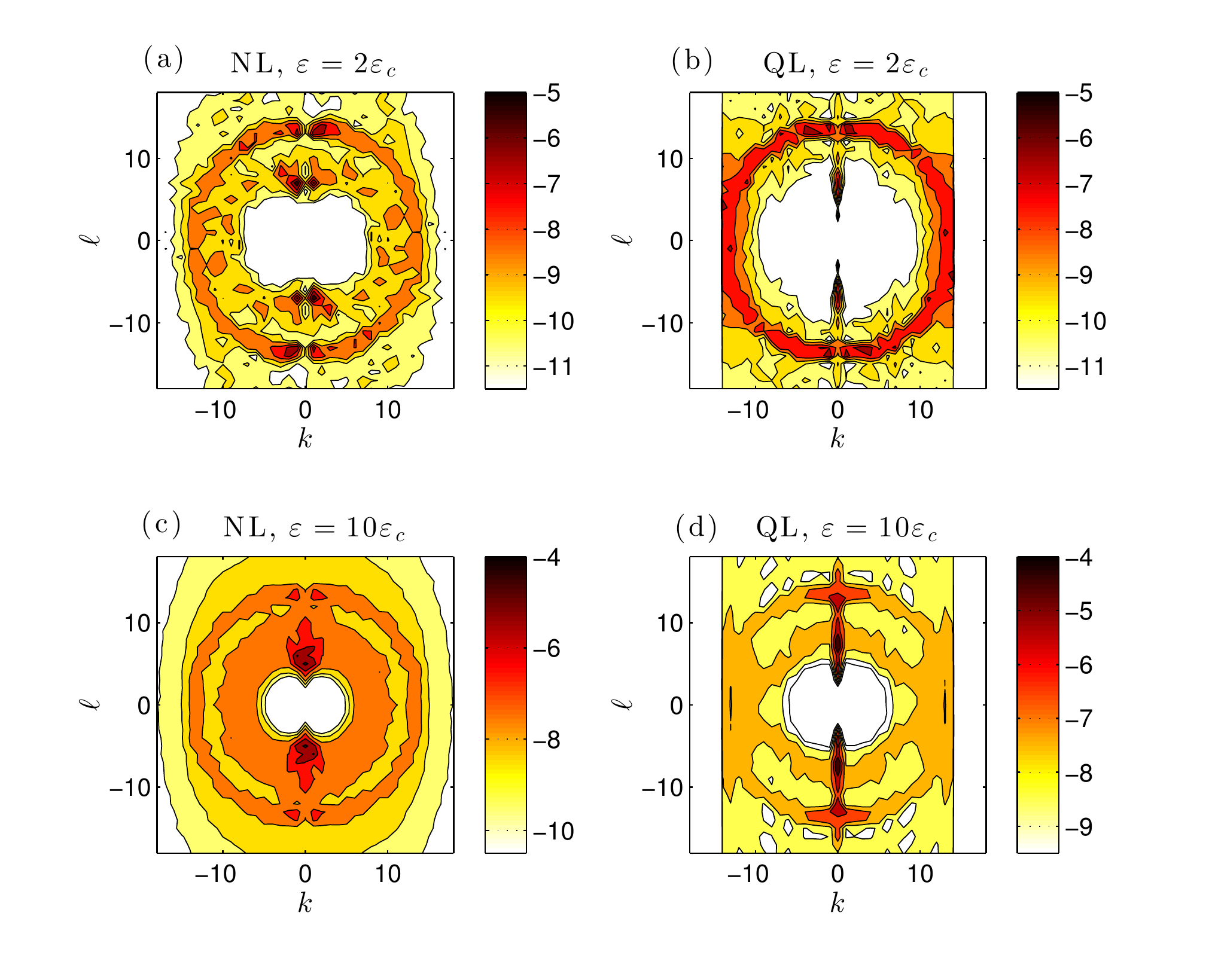}
\caption{\label{fig:spectra_NL_QL_IRFh_r_0p01_f2_f10}The statistical equilibrium enstrophy spectrum, $\log{\( \<|\tilde{\z}_{k\l}|^2\>\)}$, for NL and QL simulations under IRFn forcing  at $\varepsilon=2\varepsilon_c$ (panels (a) and (b)) and $\varepsilon=10\varepsilon_c$ (panels (c) and (d)). 
For $\varepsilon=2\varepsilon_c$ the NL simulations do not support zonal jets and  energy is seen to accumulate  
in the non-zonal structure $(|k|,|\l|)=(1,7)$. At $\varepsilon=10\varepsilon_c$,  persistent zonal jets emerge (cf.~Fig.~\ref{fig:hov_IRFh_e10ec_S3Tb}) suppressing the power in the non-zonal structures.  Parameters: $\beta=10$, $r=0.01$.}
 
%
\centering\includegraphics[width=19pc,trim=7mm 11mm 7mm 1mm,clip]{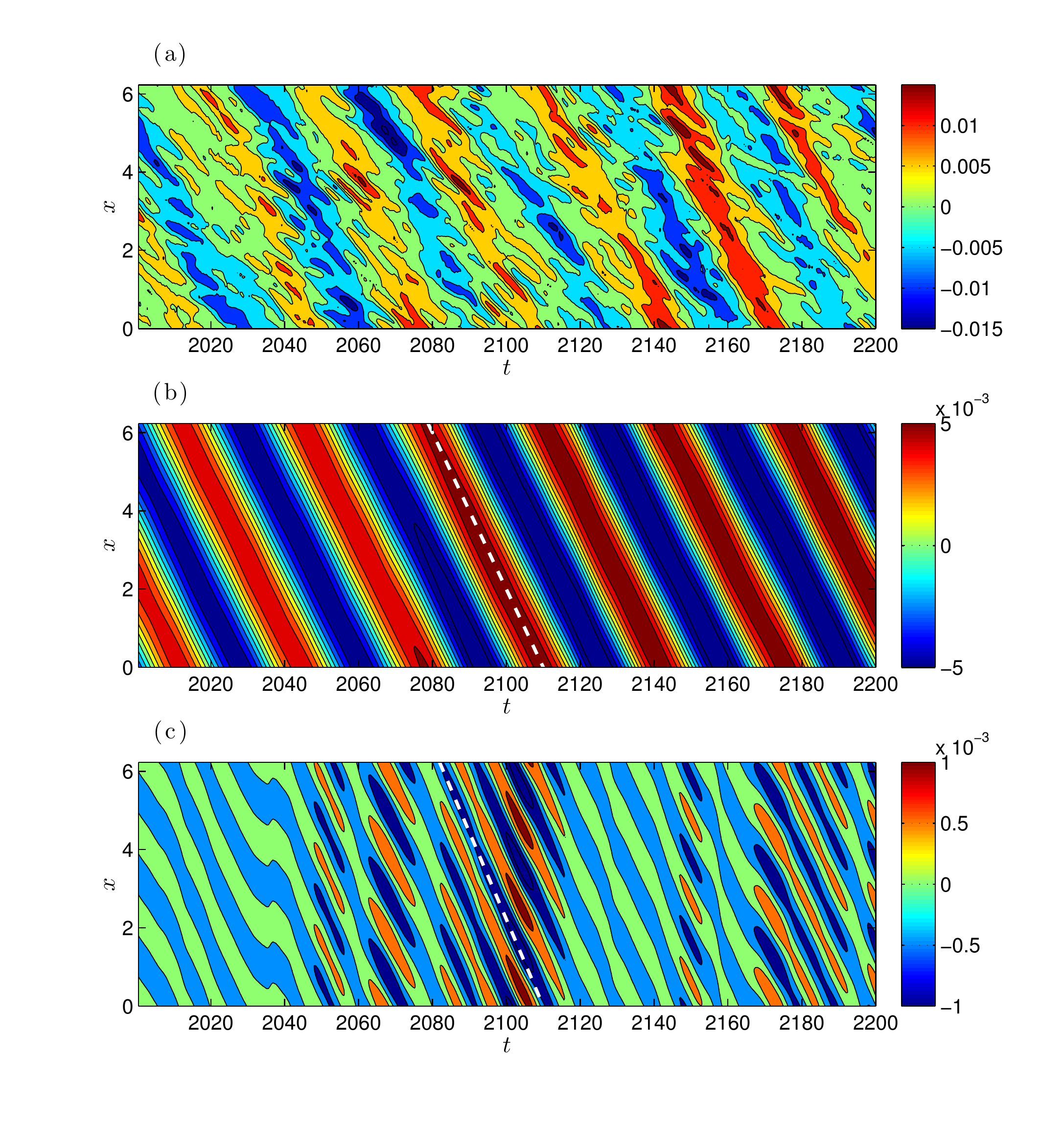}
\caption{\label{fig:nz_hov_IRFh_e2ec_r0p01} Hov\-m\"oller diagrams of the non-zonal structures supported in the NL simulation of Fig.~\ref{fig:hov_IRFh_e2ec_r0p01}. Panel (a): evolution of the  total perturbation
 streamfunction, $\psi(x,y=y_0,t)$, at latitude $y_0=\pi/4$. Panel (b): evolution of the  
 dominant $(|k|,|\l|)=(1,7)$ structure  of  $\psi(x,y=y_0,t)$ at latitude $y_0=\pi/4$. Almost half of the energy input to the system is captured and dissipated by this mode, which is phase coherent and propagates at the Rossby wave speed indicated by the dashed line. Panel (c): evolution   of  the $(|k|,|\l|)=(3,6)$ structure at the same latitude. While  this structure propagates at the Rossby wave speed it is not phase coherent.  Parameters: IRFn forcing at $\varepsilon=2\varepsilon_c$, $\beta=10$, $r=0.01$.}
\end{figure}

Despite S3T supercriticality, no persistent jets emerge in NL simulations 
with IRFn forcing in the interval  $\varepsilon_c<\varepsilon<2.8 \varepsilon_c$ 
(cf.~Fig.~\ref{fig:bifurc_NIF_IRFh_r0p01}\hyperref[fig:bifurc_NIF_IRFh_r0p01]{a}).
Comparisons of  NL, QL and S3T simulations with IRFn forcing at 
$\varepsilon=2 \varepsilon_c $ are 
shown in Fig.~\ref{fig:hov_IRFh_e2ec_r0p01}. 
Instead of zonal jets, in the NL simulation prominent non-zonal structures are seen to propagate westward at the Rossby wave phase speed.
These non-zonal structures are also evident in the concentration of power in the enstrophy spectrum at  $(|k|,|\l|)=(1,7)$ (cf.~top panels of Fig.~\ref{fig:spectra_NL_QL_IRFh_r_0p01_f2_f10}).
At this forcing amplitude these structures are essentially linear Rossby waves which, 
if stochastically forced, would be coherent only over the dissipation time scale $1/r$. Coherence  on the dissipation time scale is observed in the subdominant part of the spectrum as seen in the case of the $(3,6)$ structure in  Fig.~\ref{fig:nz_hov_IRFh_e2ec_r0p01}\hyperref[fig:nz_hov_IRFh_e2ec_r0p01]{c}.
However, the dominant $(1,7)$ structure remains coherent  over  time  periods  far exceeding the  dissipation time scale (cf. Hov\-m\"oller diagram Fig.~\ref{fig:nz_hov_IRFh_e2ec_r0p01}\hyperref[fig:nz_hov_IRFh_e2ec_r0p01]{b}).
This case represents a regime in which the flow is dominated by a single non-zonal  structure. 
Both the concentration of power  in and  the coherence of this  structure will  be addressed below.

\begin{figure}[h!]
\centering\includegraphics[width=19pc]{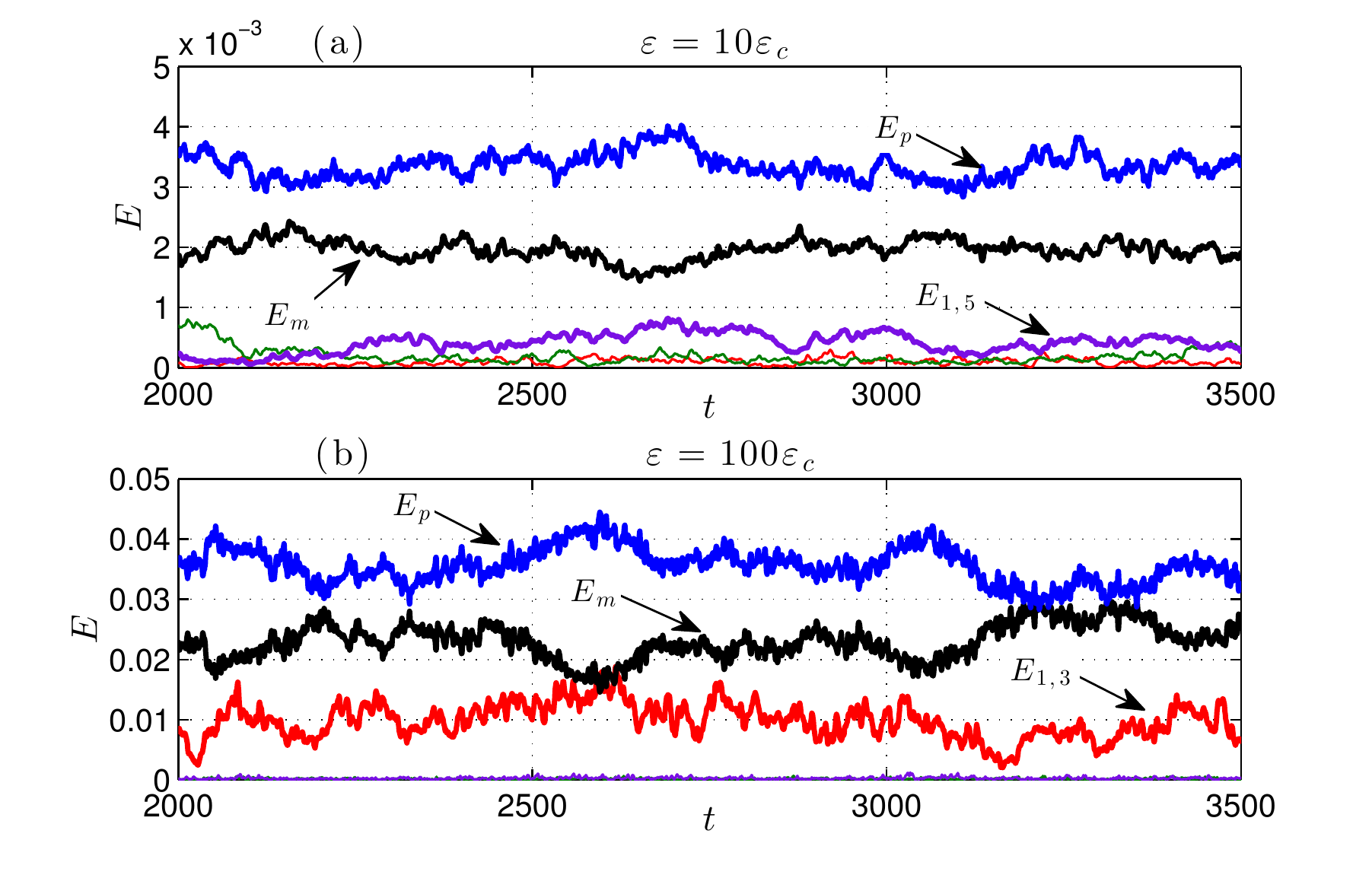}
\caption{Panel (a): Evolution of the mean flow energy, $E_m$, which is concentrated at $(0,6)$, the total eddy energy, $E_p$, and the energy of the (1,5), (1,6) and (1,7) structures for the NL simulation with IRFn forcing at $\varepsilon=10\varepsilon_c$, shown in Fig.~\ref{fig:hov_IRFh_e10ec_S3Tb}.  Panel (b): Evolution of the mean flow energy, $E_m$, the total eddy energy, $E_p$, as well as the energy of the (1,3), (1,5) and (1,6) structures for the NL simulation with IRFn forcing at $\varepsilon=100\varepsilon_c$, shown in Fig.~\ref{fig:hov_IRFh_e100ec_r0p01}. The mean flow energy is concentrated at $(0,3)$.
 In both panels the evolution of the energies is shown after statistical steady state has been reached.}
  \label{fig:Em_Eps_f10_f100}
\end{figure}

When the forcing is increased to  $\varepsilon = 10 \varepsilon_c$, a $(0,6)$ jet structure emerges, suppresses the non-zonal $(1,7)$ structure, and becomes  the dominant structure. A prominent  phase coherent non-zonal $(1,5)$ structure propagating with the Rossby wave speed is also present,  as shown in Fig.~\ref{fig:Em_Eps_f10_f100}.
 A similar regime of coexisting jets and non-zonal structures is also evident at higher supercriticalities.  An example is  the case of the equilibrium state at $\varepsilon=100 \varepsilon_c$ (cf. Fig.~\ref{fig:hov_IRFh_e100ec_r0p01}) in which the energy of the flow is shared between the (0,3) jet and the $(1,3)$ structure, as shown in  Fig.~\ref{fig:Em_Eps_f10_f100}. At this forcing level  the $(1,3)$ structure is not phase coherent, but its phase speed is still given by the Rossby wave speed. At even higher forcing similar non-zonal  structures, referred to as zonons, have been reported to coexist  with zonal jets while propagating  phase incoherently at speeds that differ substantially from the Rossby wave speed ~\citep{Sukoriansky-etal-2008}. These cases provide examples of the regime in which jets and non-zonal structures coexist. 

\begin{figure}[h!]
\centering\includegraphics[width=19pc,trim=5mm 1mm 5mm 1mm,clip]{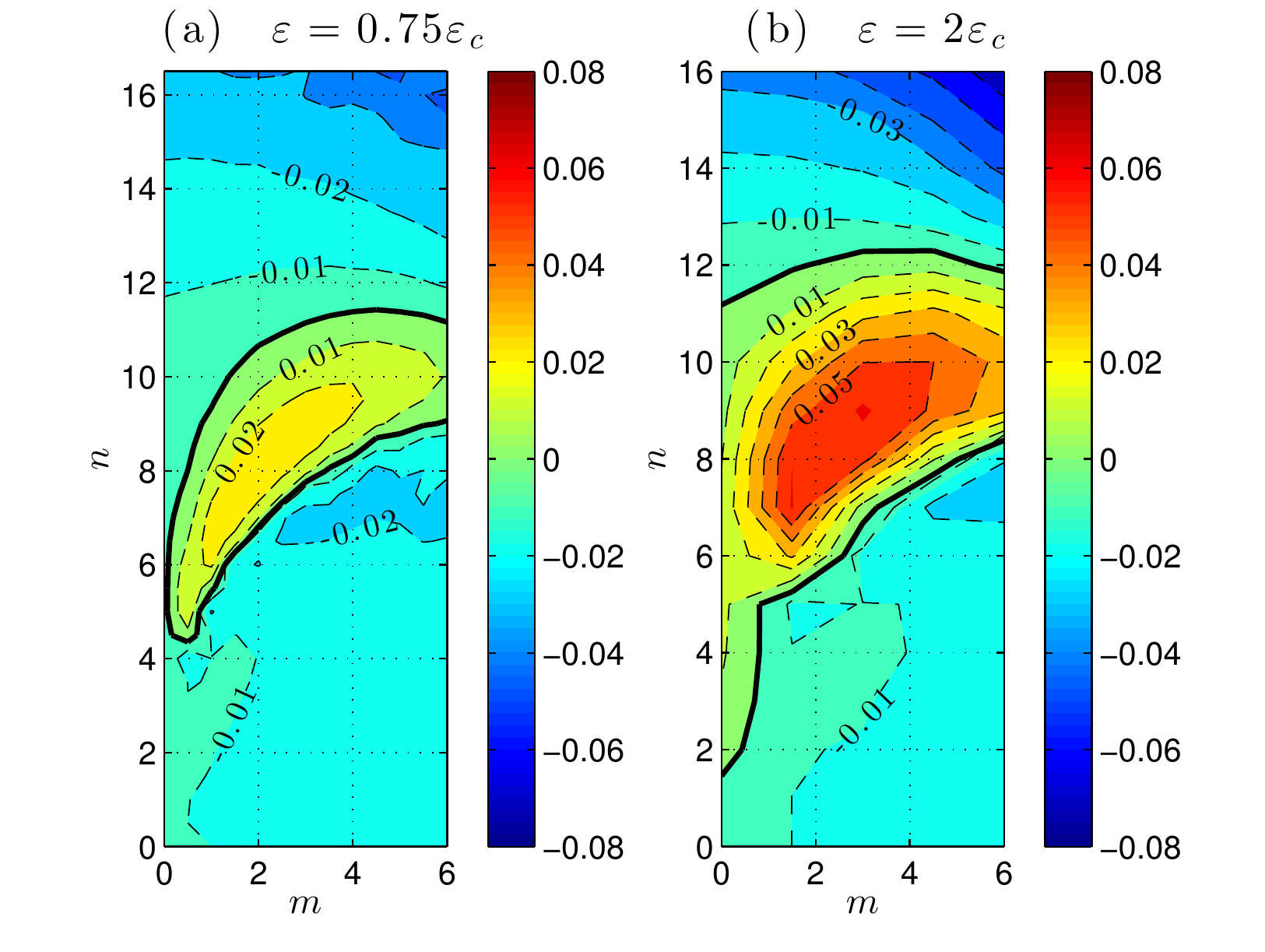}
\caption{Growth rate, $\sigma_r$,  of the  S3T non-zonal  eigenfunction, $e^{\i(mx+ny)}$, as a function of zonal  wavenumber $m$ and meridional wavenumber, $n$ for IRFn  at $\varepsilon=0.75\varepsilon_c$ (panel (a)) and $\varepsilon=2\varepsilon_c$ (panel (b)). Here $\varepsilon_c$ is the critical energy input rate for the emergence of jets. The values at the axis, $(0,n)$,  give the growth rate of  the corresponding jet perturbation. For $\varepsilon =0.75  \varepsilon_c$  the $m=0$ jet eigenfunctions are  stable but the non-zonal perturbations are unstable with  maximum instability occurring  at $(m,n)=(2,8)$. For $\varepsilon=2\varepsilon_c$ the  $m=0$ perturbations are unstable but  the non-zonal perturbations are more strongly unstable, with maximum growth at $(m,n)=(2,8)$ and  $(m,n)=(1,7)$. An NL simulation at $\varepsilon=2\varepsilon_c$ accumulated energy  at $(|k|,|\l|)=(1,7)$ (cf.~Fig.~\ref{fig:spectra_NL_QL_IRFh_r_0p01_f2_f10}) while the vorticity field showed some  accumulation at $(|k|,|\l|)=(2,8)$  (cf. Fig.~\ref{fig:NL_spectr_stability}\hyperref[fig:bifurc_assym]{f}). The stability boundary ($\s_r=0$) is marked with thick solid line. 
  For both panels $\beta=10$ and $r=0.01$. }
  \label{fig:nz_S3Tgr_IRFh_r_0p01}
\end{figure}

In order to study the dynamics of  non-zonal structures within the framework of S3T  a 
different interpretation  of the ensemble mean in the S3T formulation is required: instead of interpreting the ensemble means  as zonal means, interpret them rather as  Reynolds averages  over an intermediate time scale \citep{Bernstein-2009,Bernstein-Farrell-2010, Bakas-Ioannou-2013-prl,Bakas-Ioannou-2013-jfm}.
Analysis of S3T stability of the homogeneous equilibrium state  using this broader interpretation   (cf.~\hyperref[app:nzstability]{Appendix~D}) reveals that when the energy input rate reaches the value $\varepsilon_c$,  which is the S3T stability threshold for the emergence of zonal jets, the  state may already be unstable to non-zonal structures. This can be seen in the stability analysis shown in Fig.~\ref{fig:nz_S3Tgr_IRFh_r_0p01} which reveals that the maximum growth rate occurs at wavenumbers corresponding to non-zonal structures.
In agreement with this stability analysis, the spectrum of the NL simulation shows concentration 
of power in these most S3T unstable wavenumbers (cf.~Fig.~\ref{fig:spectra_NL_QL_IRFh_r_0p01_f2_f10}).

\clearpage

The dominance and persistence of the structures seen in these NL simulations can be understood from this stability analysis and its extension into the nonlinear regime. 
Because the stochastic forcing  is white in time, the energy injection rate is fixed and state independent and, assuming linear damping at rate $r$ dominates the dissipation, the total flow energy assumes the fixed and state independent mean value $E_m+E_p=\varepsilon/ (2 r)$. At finite amplitude the set of S3T unstable structures equilibrate to allocate among themselves most of this energy  which results in the dominance of a small subset of these structures. However, we find  that  in this competition  a specific  zonal jet structure has primacy so that even if  this structure is not the most linearly unstable it emerges as the dominant structure.

%

\begin{figure}[h!]
\centering\includegraphics[width=19pc,trim=0mm 1mm 0mm 1mm,clip]{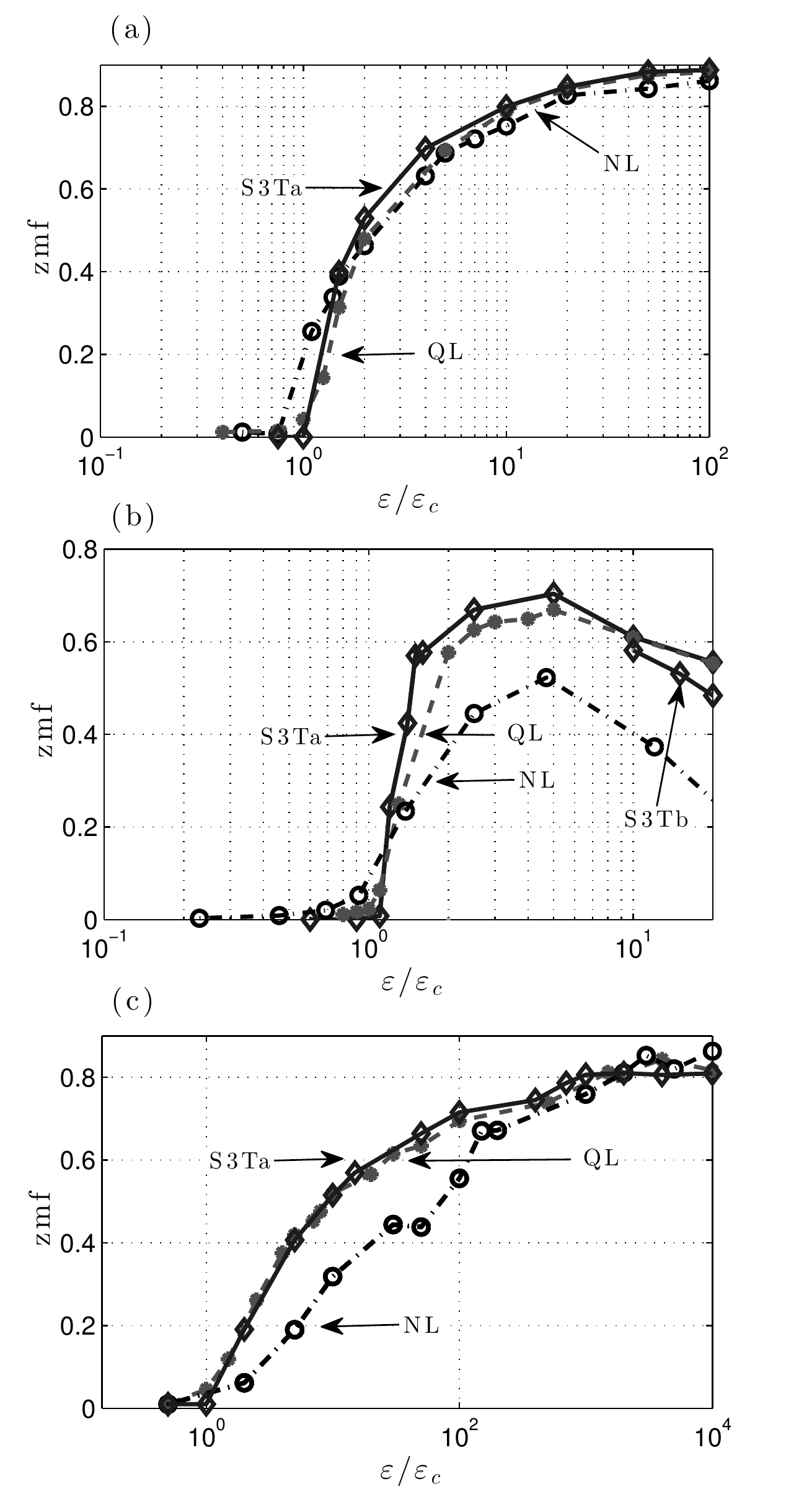}
 \vspace*{-1mm} 
 \caption{Bifurcation structure comparison for jet formation in S3T, QL, and NL with  asymmetric damping. Shown is the zmf index of jet equilibria for NIF (panel (a)), IRFn (panel (b)) and IRFw (panel (g)) as a function of the forcing amplitude $\varepsilon/\varepsilon_c$ for the NL simulation (dash-dot and circles), the QL simulation (dashed and dots) and the corresponding S3Ta simulation (solid and diamonds). Also shown in panel (b) is the zmf that is obtained  from S3T  simulations forced with the nonlinearly  modified S3Tb spectrum (calculated from  ensemble NL simulations at  $\varepsilon=20\varepsilon_c$). Parameters are $\beta=10$, $r=0.1$, $\rU=0.01$.}
  \label{fig:bifurc_assym}
\end{figure}

An attractive means for exploring the dynamics of the interaction between jets and non-zonal structures is changing  the jet  damping rate in Eq.~\eqref{eq:Ut_full} from $r$ to $\rU$  and allowing it to assume values different from the eddy damping rate, $r$, in Eq.~\eqref{eq:eddy_evol}.
In this way we can control the relative stability of jets and non-zonal structures
as well as  the finite equilibrium amplitude reached by the jet.
This asymmetric damping may be regarded as a  model for approximating
jet dynamics in a baroclinic flow in which  the upper level jet is lightly damped, while the  active baroclinic turbulence generating scales are  strongly Ekman damped. This asymmetry in the damping between upper and lower levels  contributes to making  jets in baroclinic turbulence generally stronger than jets in barotropic turbulence \citep{Farrell-Ioannou-2007-structure, Farrell-Ioannou-2008-baroclinic}.
By appropriate  choice of $r$ and $\rU$ a regime can be obtained in which the zonal jet instability appears first  as $\varepsilon$ increases. Because once jets are unstable they dominate non-zonal structures, in this regime zonal jets are the dominant  coherent structure and 
S3T analysis based on the zonal interpretation of the ensemble mean produces very good  agreement with NL. For example, a comparison of bifurcation structures  among S3T, QL and 
NL under NIF and IRFn forcing using  the asymmetric  damping  $r=0.1$ and   $\rU=0.01$ 
demonstrates that jets emerge at the same critical value in S3T, QL and NL (cf.~Figs.~\ref{fig:bifurc_assym}\hyperref[fig:bifurc_assym]{a} and \ref{fig:bifurc_assym}\hyperref[fig:bifurc_assym]{b}).
This agreement,  which has been obtained by asymmetric damping induced suppression  of the non-zonal instability up to $\varepsilon_c$, implies that in the simulations with symmetric damping the difference between the S3T prediction for the  forcing amplitude $\varepsilon$ required for the first emergence of jets and the value of $\varepsilon$ obtained for first jet appearence in NL (cf.~Figs.~\ref{fig:bifurc_NIF_IRFh_r0p01}\hyperref[fig:bifurc_NIF_IRFh_r0p01]{a} and \ref{fig:bifurc_NIF_IRFh_r0p01}\hyperref[fig:bifurc_NIF_IRFh_r0p01]{b})
can  be attributed to modification of the background spectrum by  the prior emergence  of the non-zonal structures predicted by S3T.
Once unstable, zonal structures immediately dominate non-zonal structures which explains why S3T dynamics based on the zonal mean interpretation of the ensemble mean produces accurate results for parameter values for which zonal jets are the first instability to occur.
 
\begin{figure}[h!]
\centering\includegraphics[width=19pc]{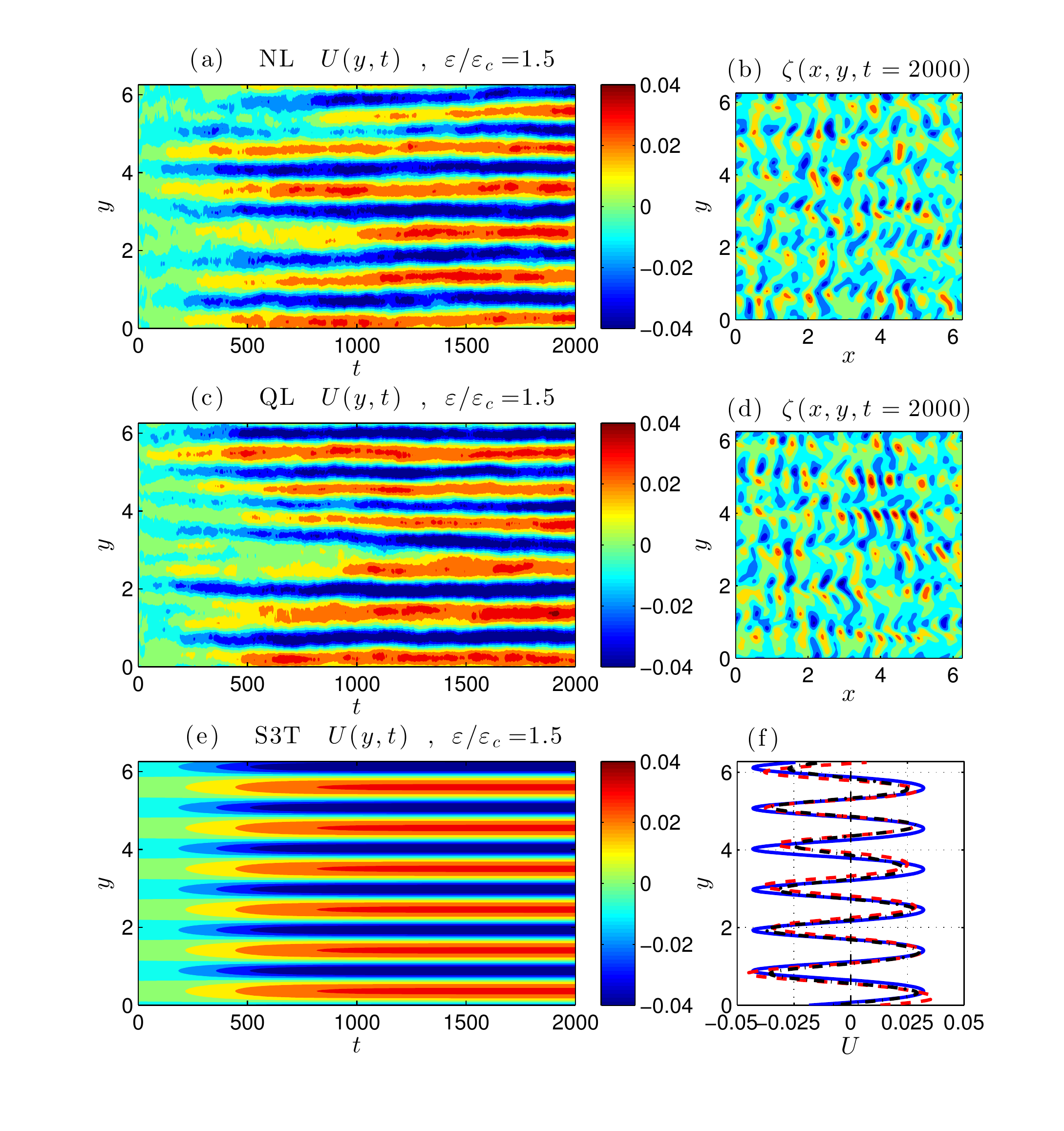}
 \caption{Hovm\"oller diagrams of  jet   emergence 
 in the NL, QL and S3T  simulations for NIF  at  $\varepsilon=1.5\varepsilon_c$ with asymmetric damping. 
 Shown is $U(y,t)$ for the NL (panel (a)), QL (panel (c)) and S3T (panel (e)) simulations  and  also characteristic snapshot of the vorticity fields at $t=2000$  for NL and QL simulations (panels (b) and (d)). Also shown are the equilibrium jets in  the NL (dash-dot), QL (dashed), and S3T (solid) simulation (panel (f)). 
 This figure shows that  S3T  predicts the structure, growth and equilibration of weakly forced jets 
 in both the  QL and NL simulations. Parameters are: $\beta=10$, $r=0.1$, $\rU=0.01$.}
 \label{fig:hov_NIF_e1p5ec_r0p1_rm0p01}
\end{figure}

A comparison of the development of  jets  in S3T, QL, and NL with this asymmetric damping and NIF forcing, shown in Fig.~\ref{fig:hov_NIF_e1p5ec_r0p1_rm0p01}, demonstrates the accuracy of the S3T predictions.
S3T stability analysis predicts that  in this case with NIF forcing maximum instability occurs at $n=6$. When these  maximally growing eigenfunctions are introduced  in the S3T system the jets grow exponentially at first at the predicted rate and then equilibrate.  
Corresponding simulations with the QL and NL dynamics  reveal nearly identical  jet growth  followed  by   finite amplitude equilibration (shown in Fig.~\ref{fig:hov_NIF_e1p5ec_r0p1_rm0p01}). Similar results are obtained with IRFn forcing. This demonstrates that the S3T dynamics comprises  both  the jet instability
mechanism  and the mechanism of  finite amplitude equilibration.    

Although no theoretical prediction of this bifurcation behavior can be made directly from  NL or QL, they both reveal the bifurcation structure obtained from the S3T analysis. 
By suppressing the peripheral complexity of non-zonal structure formation by non-zonal S3T instabilities, these simulations  allow construction of a simple model example that provides compelling evidence for identifying jet formation and equilibration in NL with the S3T theoretical framework.
Moreover, agreement among the NL, QL and S3T bifurcation diagrams shown in Figs.~\ref{fig:bifurc_assym}\hyperref[fig:bifurc_assym]{a} and \ref{fig:bifurc_assym}\hyperref[fig:bifurc_assym]{b} provides convincing evidence that turbulent cascades, which are absent in S3T or QL, are not required  for jet formation.

%

While under NIF agreement between NL and S3T  equilibrium jet amplitudes extends to all values of $\varepsilon$, under  IRFn  the NL and S3T equilibrium amplitudes diverge at larger values of $\varepsilon$ (cf.~Figs.~\ref{fig:bifurc_assym}\hyperref[fig:bifurc_assym]{a} and \ref{fig:bifurc_assym}\hyperref[fig:bifurc_assym]{b}).
This difference among NL, QL and S3T  at large $\varepsilon$ cannot be attributed to  nonlinear modification of the spectrum, which is accounted for by use of the S3Tb spectrum (cf. S3Tb response in Fig.~\ref{fig:bifurc_assym}\hyperref[fig:bifurc_assym]{b}). 
Rather, this difference is primarily due to nonlinear eddy-eddy interactions retained in NL that disrupt the up-gradient momentum transfer. This disruption is accentuated by the peculiar efficiency with which the narrow ring forcing, IRFn,  gives rise to vortices, as  can seen in  Fig.~\ref{fig:stripe_ring_Qkl}\hyperref[fig:stripe_ring_Qkl]{d-f}. The more physical distributed forcing structures do not share this property (cf.~Fig.~\ref{fig:stripe_ring_Qkl}).
We verify that the narrow ring IRFn forcing is responsible  for depressing NL equilibrium jet strength at high supercriticality by broadening the forcing distribution to assume the form IRFw (cf.~\hyperref[app:forcing]{Appendix~B} as well as Fig.~\ref{fig:stripe_ring_Qkl} for IRFn/IRFw comparison). Using IRFw while retaining other parameters as in Fig.~\ref{fig:bifurc_assym}\hyperref[fig:bifurc_assym]{b}, we obtain agreement between S3T, QL and NL simulations,
as is shown in Fig.~\ref{fig:bifurc_assym}\hyperref[fig:bifurc_assym]{c}.

\section{Identification of intermittent jets with stable S3T zonal eigenfunctions}

For subcritical forcing S3T predicts a stable  homogeneous statistical equilibrium and a set of 
eigenfunctions that govern the decay of perturbations to this equilibrium. 
We wish to show that these eigenfunctions are excited in NL by fluctuations in the turbulence
and that this excitation gives rise  in NL simulations to the formation of intermittent  jets with the form of these eigenfunctions.

\begin{figure}[h!]
\centering\includegraphics[width=19pc]{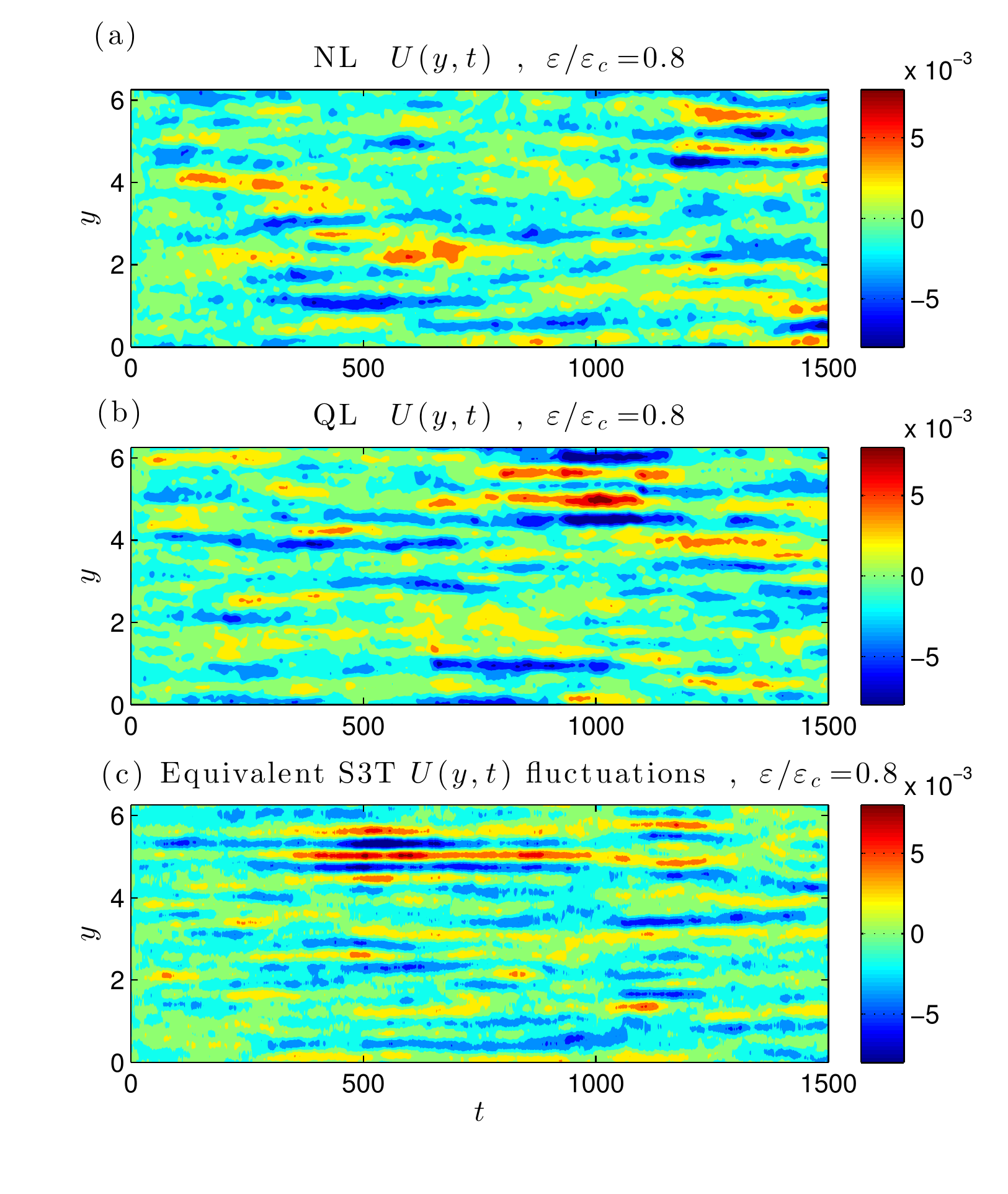}
\vspace{-5mm}\caption{ Hovm\"oller diagrams of intermittent jet structure in NL and QL simulations at subcritical forcing $\varepsilon=0.8 \varepsilon_c$. Shown are $U(y,t)$ for NL (panel (a)) and QL (panel (b)) simulations and the $U(y,t)$  that results from  random excitation of the S3T  damped modes (panel (c)). These plots were obtained using  IRFn forcing  with $r=0.1$, $\rU=0.01$. This figure shows that  the manifold of S3T damped modes are revealed by being excited in the fluctuating NL and QL simulations. Planetary vorticity gradient: $\beta=10$. } 
  \label{fig:hov_IRFh_e0p8ec_r0p1_rm0p01}
\end{figure}

As an  example, consider the simulation with asymmetric damping and IRFn subcritical forcing shown in Fig.~\ref{fig:hov_IRFh_e0p8ec_r0p1_rm0p01}. For these parameters the least damped eigenfunctions are zonal jets and  confirmation that the intermittent jets in NL, shown in the top panel of Fig.~\ref{fig:hov_IRFh_e0p8ec_r0p1_rm0p01}, are consistent 
with turbulence fluctuations exciting
the S3T damped modes is given in the bottom panel of 
Fig.~\ref{fig:hov_IRFh_e0p8ec_r0p1_rm0p01} where the intermittent jets resulting  
from stochastic forcing of the S3T modes themselves are shown. 
This diagram was obtained by plotting $U(y,t) = \real{ \[ \sum_{n=1}^N \alpha_n(t) e^{\i n y} \]}$, with $\alpha_n$ independent red noise processes, associated with the damping rates, $|\sigma(n)|$,  of  the first $N=15$ least damped  S3T modes.  These $\alpha_n$ are obtained from  the Langevin equation, $\df \alpha_n \big/ \df t = {\sigma(n)}\,\alpha_n + \xi(t)$, with $\xi(t)$ a $\delta$-correlated complex valued random variable.

The fluctuation-free S3T simulations reveal persistent jet structure only coincident with the inception of the S3T instability, which occurs only for supercritical forcing. However, in   QL and NL  simulations fluctuations excite the damped manifold of modes predicted by the S3T analysis to exist at subcritical forcing amplitudes. This observation confirms the reality of the   manifold of S3T stable modes. 

In NL and QL simulations these stable modes predicted by S3T are increasingly excited as the critical bifurcation point in parameter space is approached, because their damping rate vanishes at the bifurcation. The associated increase in zonal mean flow energy on approach to  the bifurcation point  obscures the exact location of the bifurcation point  in NL and QL simulations compared to the fluctuation-free S3T simulations for which the bifurcation is exactly coincident with the inception of the S3T instability  (i.e. Fig.~\ref{fig:bifurc_assym}).

\section{Verification in NL of the multiple jet equilibria predicted  by S3T}

\begin{figure}[t]
\centering\includegraphics[width=19pc]{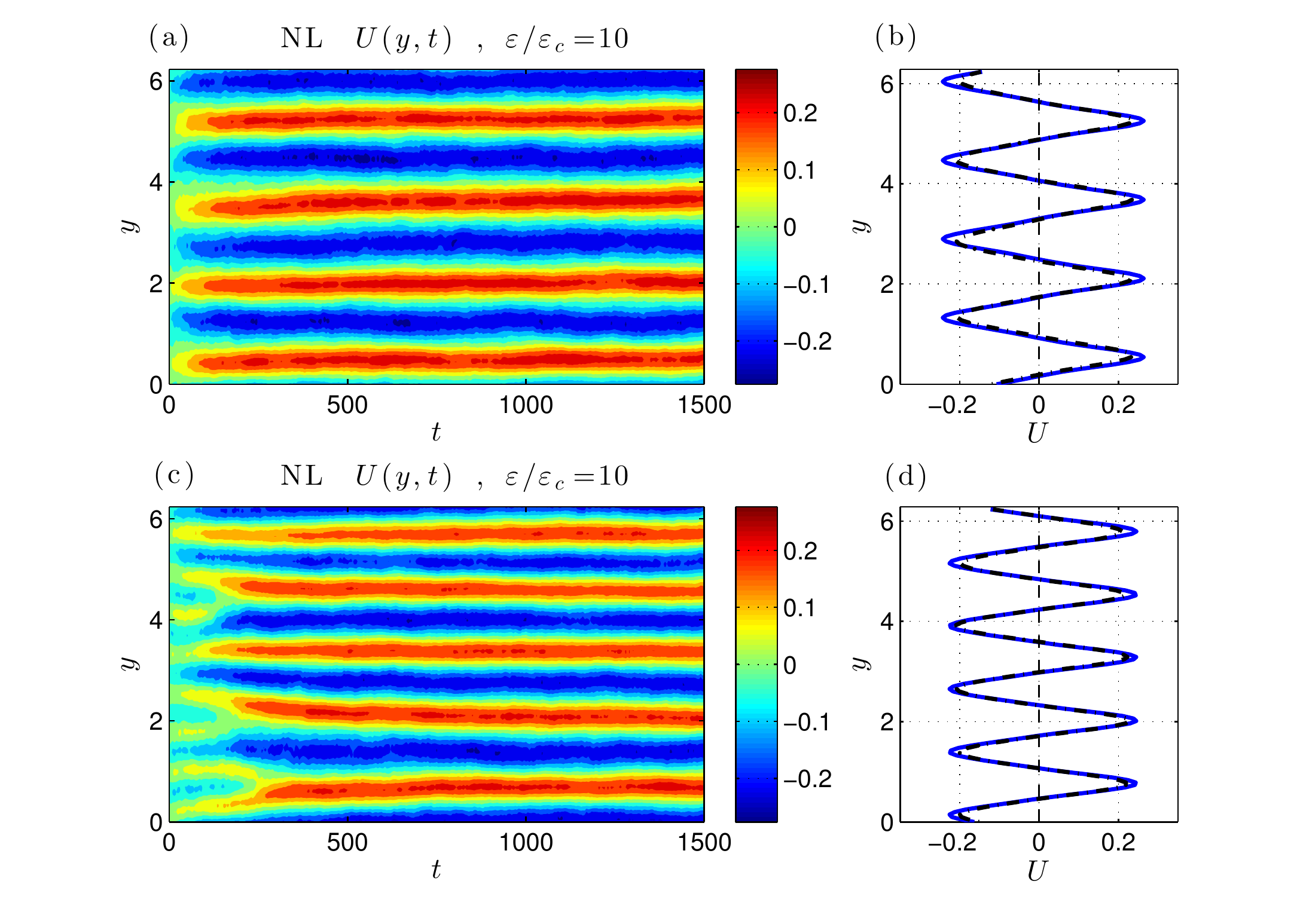}
\caption{Realizations in NL simulations of multiple equilibria predicted by S3T.  Show are Hovm\"oller diagrams  of NL simulations showing  the equilibrium with 4 jets (panel (a)) and with 5 jets (panel (c)). Also shown is comparison of the S3T equilibrium jets (solid) with the average jets obtained from the NL simulation (dashed) for the two equilibria (panels (b) and (d)). Parameters: NIF forcing  at $\varepsilon=10 \varepsilon_c$, $r=0.1$, $\rU=0.01$ and $\beta=10$.}
  \label{fig:multiple_eq}
\end{figure}

As is commonly found in nonlinear systems, the finite amplitude equilibria predicted by S3T are not necessarily unique and multiple equilibria can occur for the same parameters.   
S3T provides a theoretical framework for studying  these multiple equilibria, their stability and bifurcation structure. An example of two such S3T equilibria are shown in Fig.~\ref{fig:multiple_eq}  together with their associated NL simulations.
As the parameters change these equilibria may cease to exist or become S3T unstable. 
Similar multiple equilibria have been found in  S3T studies of barotropic beta-plane turbulence \citep{Farrell-Ioannou-2003-structural,Farrell-Ioannou-2007-structure,Parker-Krommes-2013-generation} and in S3T studies of baroclinic turbulence~\citep{Farrell-Ioannou-2008-baroclinic,Farrell-Ioannou-2009-closure} and the hypothesis has been advanced that the existence of such  multiple jet equilibria may underlie the abrupt transitions found in  the record of Earth's climate~\citep{Farrell-Ioannou-2003-structural, Wunsch-2003}.

\section{Conclusions}

In this work  predictions of  S3T for  jet formation and equilibration
in barotropic beta-plane turbulence were critically compared with results obtained using QL and NL simulations.
The qualitative bifurcation structure  predicted by S3T for emergence of zonal jets from a homogeneous turbulent state was confirmed  by both  the QL and NL simulations. Moreover, the finite amplitude  equilibrium jets in NL and QL  simulations were found to be as predicted by the fixed point solutions of  S3T. 
Differences in jet formation bifurcation parameter values between NL and QL/S3T were reconciled by taking account of the fact that the spectrum of turbulence is substantially modified in NL.  Remarkably, the modification of the spectrum in NL could be traced in large part to emergence of non-zonal  structures through S3T instability. 
When account is taken of the modification of the turbulent spectrum resulting substantially from these 
non-zonal  structures, S3T also provides quantitative agreement with the 
threshold values for the emergence of jets in NL. 
The influence of the background eddy spectrum on the S3T dynamics was 
found to be immediate, in the sense that  in spin-up simulations jets 
emerge in accordance with the instability calculated on the temporally developing spectrum.   
The fact that jets are prominent in observations is consistent with 
the robust result that when a jet structure emerges it has primacy 
over the non-zonal structures, so that even if the jet eigenfunction  
is not the most linearly S3T unstable eigenfunction, the jet still emerges at finite amplitude as the dominant structure.

%


These results confirm that jet emergence and equilibration in barotropic beta-plane turbulence results from the cooperative quasi-linear mean flow/eddy instability that is  predicted by S3T. 
These results also establish that turbulent cascades are not required for the formation of zonal jets in beta-plane turbulence. Moreover, the physical reality of the manifold of  stable modes arising  from cooperative interaction between incoherent turbulence and coherent jets, which is predicted by S3T,  was verified in this work by relating  observations of intermittent jets in NL and QL to stochastic excitation by the turbulence of  this manifold of  stable S3T modes.

S3T provides an autonomous, deterministic nonlinear closure of turbulence dynamics at second order that  provides an attractive vehicle for further investigation of the dynamics of  turbulent flows.

%

%

\begin{acknowledgment}
The authors would like to  acknowledge discussions with N. Bakas, F. Bouchet, K. Srinivasan and W. Young. Navid Constantinou acknowledges the support of  the Alexander S. Onassis Public Benefit Foundation. Brian Farrell was supported by  NSF AGS-1246929 and ATM-0736022. Brian Farrell and Petros Ioannou acknowledge the  hospitality during June 2012 of the  Aspen Center for Physics (supported by NSF under grant No.~1066293) where part of this paper was written. Petros Ioannou acknowledges  the generous support of the John S. Latsis Foundation under  ``Research Projects 2011".
\end{acknowledgment}

\ifthenelse{\boolean{dc}}
{}
{\clearpage}

\begin{appendix}[A]
\addcontentsline{toc}{section}{\protect\numberline{}Appendices}
\section*{\begin{center}Numerical details and parameters\end{center}\label{app:A}}

%
Both the nonlinear (NL) simulations of Eq.~\eqref{eq:q} and the quasi-linear (QL) simulations of Eqs.~\eqref{eq:ql_eq} were carried out with a pseudospectral Fourier code. The maximum resolved wavenumbers were  $k_{\max}=N_x/2$ and  $\ell_{\max}=N_y/2$ and the maximum resolved total wavenumber $K_{\max}=\(k_{\max}^2+\l_{\max}^2\)^{1/2}$. For the time integration a  fourth order Runge-Kutta method (RK4) was used together with a Godunov step for integrating the stochastic forcing. In all calculations hyperviscosity was added for numerical stability with coefficient $\nu_4 = 0.5\big/ (K^4_{\max}\, \d t)$, where $\d t$ is the time step. In all calculations $\d t=2.5\times 10^{-3}$ and $N_x=N_y=256$, which imply $\nu_4 = 1.86\times 10^{-7}$. 



\end{appendix}

\ifthenelse{\boolean{dc}}
{}
{\clearpage}

\begin{appendix}[B]
\section*{\begin{center}The stochastic forcing structure and its associated power spectrum\end{center}}\label{app:forcing}

The stochastic forcing at point $(x_i,y_j)$ is
\be
F(x_i,y_j,t) = \,\real{\[\sum_{k=1}^{N_k} \sum_{{p}=1}^{N_y}\[\bit \F_k\]_{j {p}}\xi_{k{p}}(t)\,e^{\i k x_i}\]}~,
\label{eq:forcing_realization}
\ee
in which $\xi_{k{{p}}}$ are temporally $\delta$-correlated and independent  and satisfy: 
$\<\xi_{k{{p}}}(t)\bit \>=0$, $\<\xi_{k{{p}}}(t)\,\xi_{mn}^*(t')\bit \> = \d_{km}\d_{{{p}} n}\,\d(t-t')$.
The stochastic forcing is correlated in $y$  by the columns of the matrix  $\F_k$. For the  non-isotropic forcing (NIF) this meridional structure, in a periodic domain with period $2\pi$ in $y$,  is specified by:
\ifthenelse{\boolean{dc}}
{\begin{align}
\[\bit \F_k\]_{j{p}} =  c_k\[e^{-(y_j-y_{{p}})^2/(2s^2)} \right.+& \left.e^{-(y_j-2 \pi -y_{{p}})^2/(2s^2)}\right.+\nonumber\\
&~~\left. + e^{-(y_j+2 \pi-y_p)^2/(2s^2)}\]~.\label{eq:Fk_wave}
\end{align}
}{\be
\[\bit \F_k\]_{j{{p}}} =  c_k\[e^{-(y_j-y_{{p}})^2/(2s^2)} + e^{-(y_j-2 \pi -y_{{p}})^2/(2s^2)} + e^{-(y_j+2 \pi-y_{{p}})^2/(2s^2)}\]~.\label{eq:Fk_wave}
\ee
}
We choose   $s=0.2/\sqrt{2}$ and  force  zonal components $k=1,\dots,14$ .
Because the stochastic forcing is $\delta$-correlated in time,
the energy input rate,  given by:
\be
\mathcal{E} = - \int \frac{\df x}{L_x}\frac{\df y}{L_y} \,\psi' F~=\sum_{k=1}^{N_k} -\frac1{4 N_y} \Tr\(\DDel_k^{-1}\F_k^{\vphantom{\dag}}\F_k^\dag\)~,\label{eq:e_k_injection} 
\ee
does not depend on the  state of the system and can be independently specified. 
The normalization constant, $c_k$, in (\ref{eq:Fk_wave}) is chosen so  that each  $k$  is excited equally and one unit of energy is injected in total. It follows that the  total energy input rate in the NL,  QL or  S3T simulations is given by $\varepsilon$.

The isotropic ring forcing  is specified by:
\be
\[\bit \F_k\]_{jp} = c \,w_{kp} (K) \,e^{\i \l(p) \, y_j}\label{eq:Fk_isotropic}~,
\ee
with $\l(p) = (p-1)-N_y/2$ and $K=\sqrt{k^2 +\l^2}$. Meridional wavenumber $\l$ extends from  $-N_y/2$ to $N_y/2-1$ because only these wavenumbers are resolved  when  $N_y$ points in the meridional direction are retained (for $N_y$ even). For IRFn:
$w_{kp}=1$ for $|K -K_f| \le  \d k_f $ and zero otherwise, with
$K_f=14$ and $\d k_f = 1$.  For IRFw: $w_{k{{p}}} = c\,\exp{\[-{\(K-K_f\)^2}/({2\,\d k_f^2})\]} $ with  
$K_f=14 $ and $\d k_f = 8/\sqrt{2}$. The normalization constant, $c$, is chosen for both cases so that the total energy input rate is unity. 
IRFn and IRFw  are both spatially homogeneous and nearly isotropic in a finite doubly periodic domain. They approach  exact isotropy as the domain size increases.

The spatial covariance of the forcing, $Q(x_a,x_b,y_a,y_b)= {\<\bit  F(x_a,y_a,t) F(x_b,y_b,t) \>}$, being homogeneous in both $x$ and $y$, depends only on $x_a-x_b$ and $y_a-y_b$, and has Fourier expansion:
\ifthenelse{\boolean{dc}}
{\begin{align}
Q(x_a-&x_b,y_a-y_b)= \nonumber\\
&=\real\left\{ \sum_{k=1}^{N_k} \sum_{\l=-N_y/2}^{N_y/2-1}  \hat{Q}_{k\l} \,e^{\i \[k(x_a-x_b) + \l(y_a-y_b)\bit \]}\right\}~.\label{eq:Q_ab_to_Q_kl}
\end{align}
}
{\begin{align}
Q(x_a-x_b,y_a-y_b)&= \real\left\{ \sum_{k=1}^{N_k} \sum_{\l=-N_y/2}^{N_y/2-1}  \hat{Q}_{k\l} \,e^{\i \[k(x_a-x_b) + \l(y_a-y_b)\bit \]}\right\}~.\label{eq:Q_ab_to_Q_kl}
\end{align}
}
The $\hat{Q}_{k\l}$  are  the spatial power spectrum of the stochastic forcing. 
%
%
Fourier coefficients of the forcing covariance for only positive values of zonal wavenumbers, $k$, (cf.~Eq.~\eqref{eq:Q_ab_to_Q_kl}) can be related to Fourier expansions in both positive and negative zonal wavenumbers,
\ifthenelse{\boolean{dc}}
{\begin{align}
Q(x_a-&x_b,y_a-y_b)= \nonumber\\
&= \sum_{\substack{k=-N_k\\k\ne0}}^{N_k} \sum_{\l=-N_y/2}^{N_y/2-1}  \tilde{Q}_{k\l} \,e^{\i \[k(x_a-x_b) + \l(y_a-y_b)\bit \]}~,\label{eq:Q_ab_to_Q_kl_posneg}
\end{align}}
{
\begin{align}
Q(x_a-&x_b,y_a-y_b)= \sum_{\substack{k=-N_k\\k\ne0}}^{N_k}  \sum_{\l=-N_y/2}^{N_y/2-1}    \tilde{Q}_{k\l} \,e^{\i \[k(x_a-x_b) + \l(y_a-y_b)\bit \]}~,\label{eq:Q_ab_to_Q_kl_posneg}
\end{align}
}
through $\tilde{Q}_{k,\l} = \hat{Q}_{k,\l}\big/2$ and $\tilde{Q}_{-k,\l} = \hat{Q}_{k,-\l}\big/2$ for $k>0$. In the derivation of these relations the symmetry of the forcing covariance under exchange of the two points is used. 

\end{appendix}

\ifthenelse{\boolean{dc}}
{}
{\clearpage}

\def\s{\sigma}
\def\dCR{\d\hat{\C}_R}
\def\dCI{\d\hat{\C}_I}
\def\dAI{\d\hat{\A}_I}
\def\dUh{\d\hat{\Uv}}
\def\tzet{\d\tilde{\Uv}}
\def\txi{\tilde{\xi}}
\def\hA{\d\hat{\A}}
\def\OM{\d\hat{\boldsymbol{\Gamma}}}
\def\homp{\OM^{(+)}_{k}}
\def\homm{\OM^{(-)}_{k}}
\def\homp{\OM_{k,\textrm{P}}}
\def\homm{\OM_{k,\textrm{M}}}
\newcommand{\zetav}{\d\hat{\U}}

\begin{appendix}[C]
\section*{\begin{center}Determining the S3T stability of the homogeneous state \end{center}}\label{app:stability}

Equations~\eqref{eq:ssst_eq_pert_dt} determine the S3T stability of the equilibrium state.  Because  of the presence of the  imaginary part in Eq.~\eqref{eq:ssst_eq_pertU_dt} in order to proceed with eigenanalysis of this system we need to treat the real and imaginary part of the covariances as independent variables. Writing the covariances as  $\d\C_k = \d\C_{k,R}+\i\,\d\C_{k,I}$  and $\C^E_k =\C^E_{k,R}+\i\,\C^E_{k,I}$,
and  the operators as $\d\A_k=-\i k\[ \d\U -(\d\U)_{yy}\,\DDel^{-1}_k\] = \i\,\d\A_{k,I}$ and   $\A^E_k =\A^E_{k,R}+\i\,\A^E_{k,I}$ we  obtain the 
real coefficient system: 
\ifthenelse{\boolean{dc}}{\begin{subequations}
\begin{align}
\partial_t\,\d\Uv = &\sum_{k=1}^{N_k} \frac{1}{2}\vecd{\(\bit - k\DDel^{-1}_k \d{\C}_{k,I}\)} - \rU \,\d\Uv~,\label{eq:hzet_nosigma}\\
\partial_t\,\d\C_R = &\A^E_{R} \, \d\C_R +\d\C_R \(\A^E_{R}\)^\transp - \A^E_{I} \,\d\C_I + \nonumber\\
&\qquad+\d\C_I \(\A^E_I\)^\transp -\d\A_I\,\C^E_I +  \C^E_I\,\d\A_I^\transp~,\label{eq:dCR_nosigma}\\
\partial_t\, \d\C_I = &\A^E_{I} \,\d\C_R  -\d\C_R \(\A^E_{I}\)^\transp+\A^E_{R} \,\d\C_I +\nonumber\\
&\quad+ \d\C_I\(\A^E_R\)^\transp + \d\A_I\,\C^E_R - \C^E_R\,\d\A_I^\transp~,\label{eq:dCI_nosigma} 
\end{align}
\label{eq:hats_partialt}\end{subequations}
}{\begin{subequations}
\begin{align}
\partial_t\,\d\Uv = &\sum_{k=1}^{N_k} \frac{1}{2}\vecd{\(\bit - k\DDel^{-1}_k \d{\C}_{k,I}\)} - \rU \,\d\Uv~,\label{eq:hzet_nosigma}\\
\partial_t\,\d\C_R = &\A^E_{R} \, \d\C_R +\d\C_R \(\A^E_{R}\)^\transp - \A^E_{I} \,\d\C_I + \d\C_I \(\A^E_I\)^\transp -\d\A_I\,\C^E_I +  \C^E_I\,\d\A_I^\transp~,\label{eq:dCR_nosigma}\\
\partial_t\, \d\C_I = &\A^E_{I} \,\d\C_R  -\d\C_R \(\A^E_{I}\)^\transp+\A^E_{R} \,\d\C_I + \d\C_I\(\A^E_R\)^\transp + \d\A_I\,\C^E_R - \C^E_R\,\d\A_I^\transp~,\label{eq:dCI_nosigma} 
\end{align}
\label{eq:hats_partialt}\end{subequations}}
in which the subscript $k$ in all the variables in Eqs.~\eqref{eq:dCR_nosigma} and \eqref{eq:dCI_nosigma} has been omitted. In Eqs.~\eqref{eq:hats_partialt} the coefficient of linear damping of the mean flow, $\rU$, may differ from the coefficient of linear damping of the non-zonal perturbations, $r$ (cf. section~\ref{sec:nonzonal}). 
The asymptotic stability of  Eqs.~\eqref{eq:hats_partialt} is determined by assuming solutions of the form $(\d\hat{\Uv},\d\hat{\C}_{k,R},\d\hat{\C}_{k,I})\,e^{\sigma t}\quad\text{for }k=1,\dots,N_k$, with  $\d\A_{k,I} = \d\hat{\A}_{k,I}\,e^{\sigma t}$ and  by determining the eigenvalues,  $\sigma$, and the eigenfunctions of the system:
\ifthenelse{\boolean{dc}}
{\begin{subequations}
\begin{align}
\sigma\,\dUh = &\sum_{k=1}^{N_k} \frac{1}{2}\vecd{\(\bit -k\DDel^{-1}_k \d\hat{\C}_{k,I}\)} - \rU \,\dUh~,\label{eq:hzet}\\
\sigma\,\dCR = &\A^E_{R} \, \dCR +\dCR \(\A^E_{R}\)^\transp - \A^E_{I} \,\dCI+\nonumber\\
&\qquad + \dCI \(\A^E_I\)^\transp -\dAI\,\C^E_I +  \C^E_I\,\dAI^\transp~,\label{eq:dCR_sigma}\\
\sigma\, \dCI = &\A^E_{I} \,\dCR  -\dCR \(\A^E_{I}\)^\transp+\A^E_{R} \,\dCI + \nonumber\\
&\qquad+ \dCI\(\A^E_R\)^\transp + \dAI\,\C^E_R - \C^E_R\,\dAI^\transp~.\label{eq:dCI_sigma} 
\end{align}
\label{eq:hats}\end{subequations}}
{\begin{subequations}
\begin{align}
\sigma\,\dUh &= \sum_{k=1}^{N_k} \frac{1}{2}\vecd{\(\bit -k\DDel^{-1}_k \d\hat{\C}_{k,I}\)} - \rU \,\dUh~,\label{eq:hzet}\\
\sigma\,\dCR &= \A^E_{R} \, \dCR +\dCR \(\A^E_{R}\)^\transp - \A^E_{I} \,\dCI + \dCI \(\A^E_I\)^\transp -\dAI\,\C^E_I +  \C^E_I\,\dAI^\transp~,\label{eq:dCR_sigma}\\
\sigma\, \dCI &= \A^E_{I} \,\dCR  -\dCR \(\A^E_{I}\)^\transp+ \A^E_{R} \,\dCI + \dCI\(\A^E_R\)^\transp + \dAI\,\C^E_R - \C^E_R\,\dAI^\transp~.\label{eq:dCI_sigma} 
\end{align}
\label{eq:hats}\end{subequations}}
 In most cases direct eigenanalysis of this system is computationally prohibitive  because it involves eigenanalysis of matrices of dimension $(2 N_k N_y^2 +N_y) \times (2 N_k N_y^2 +N_y) $ if  $N_y$ grid points are used to approximate the functions and $N_k$ zonal wavenumbers are forced.
In this section we describe an efficient iterative method that can produce solutions to this stability problem for large $N_y$. The method is a generalization of the
adiabatic approximation used in earlier  studies \citep{Farrell-Ioannou-2003-structural,Farrell-Ioannou-2007-structure,Bakas-Ioannou-2011}.

When \eqref{eq:hats} has eigenvalues  with  $\real(\sigma)>0 $ the equilibrium is S3T unstable. When  $\sigma$ is complex  the eigenfunctions
$\d\hat{\Uv}$, $\d\hat{\C}_{k,R}$, $\d\hat{\C}_{k,I}$  and  $\d\hat{\A}_{k,I}$ will be complex.
Realizable, Hermitian,  solutions can then be formed by superposing  the complex conjugate eigenfunction.
Note that the  covariances  are required to be Hermitian but need not be positive definite (for a discussion of eigenvalue problems involving covariances cf. \cite{Farrell-Ioannou-2002-perturbation-II}).

Because of the periodic boundary conditions the mean flow eigenfunctions  $\dUh$  are in general a superposition of harmonics. 
However,  here we are treating the stability of the homogeneous equilibrium and the  eigenfunctions  can be shown to be
single harmonics, $\dUh_n = e^{\i n y}$, and  Eq.~\eqref{eq:hzet} becomes:
\be
\sigma \, e^{\i n y}= \sum_{k=1}^{N_k} \frac{1}{2}\vecd{\(\bit - k\DDel^{-1}_k \d\hat{\C}_{k,I}\)} - \rU \,e^{\i n y}~.\label{eq:dU_sigma_einy}
\ee
 The number of unstable jets, if the equilibrium is unstable, is $n$.   Equation~\eqref{eq:dU_sigma_einy} can be regarded as an equation for $\sigma$  given that   $\d\hat{\C}_{k,I}$ satisfies the coupled equations Eqs.~\eqref{eq:dCR_sigma}~and \eqref{eq:dCI_sigma}
 and is therefore a function of $\sigma$ and $n$.  Having transformed Eq.~\eqref{eq:dU_sigma_einy}  into an equation for $\sigma$ for a given $n$ the eigenvalues can be determined by iteration.


It is advantageous to solve   Eqs.~\eqref{eq:dCR_sigma}~and \eqref{eq:dCI_sigma} 
for the variables $\homp =\d\hat{\C}_{k,R} +\i\,\d\hat{\C}_{k,I}$, $\homm=\d\hat{\C}_{k,R} -\i\,\d\hat{\C}_{k,I}$, which satisfy the decoupled Sylvester equations:\def\dAI{\d\hat{\A}_{n,I}}
\ifthenelse{\boolean{dc}}
{\begin{subequations}
\begin{align}
0 &= (\A^E-\sigma\I)\,\homp + \homp \(\A^E\)^{\dag}  + \nonumber\\
&\qquad\qquad\quad+\(\i\,\dAI\)\C^E + \C^E  \(-\i\,\dAI^{\transp}\)~,\label{eq:homp}\\
0 &=  \[(\A^E)^*-\sigma\I\]\,\homm + \homm \[(\A^E)^*\]^\dag + \nonumber\\
&\qquad\qquad\quad+\(-\i\,\dAI\)\(\C^E\)^* + \(\C^E\)^*  \(\i\,\dAI^\transp\)~.\label{eq:homm}
\end{align}
\end{subequations}}
{\begin{subequations}
\begin{align}
0 &= (\A^E-\sigma\I)\,\homp + \homp \(\A^E\)^{\dag}  + \(\i\,\dAI\)\C^E + \C^E  \(-\i\,\dAI^{\transp}\)~,\label{eq:homp}\\
0 &=  \[(\A^E)^*-\sigma\I\]\,\homm + \homm \[(\A^E)^*\]^\dag + \(-\i\,\dAI\)\(\C^E\)^* + \(\C^E\)^*  \(\i\,\dAI^\transp\)~.\label{eq:homm}
\end{align}\label{eq:2sylvester}
\end{subequations}}
(N.B. If the  eigenvectors  $\d\hat{\C}_{k,R}$ and $\d\hat{\C}_{k,I}$  are complex, despite the  notation,  then  $\real(\homp)\ne \d\hat{\C}_{k,R}$ and $\imag(\homp)\ne \d\hat{\C}_{k,I}$.)

\end{appendix}



%
%
%
%
%
\ifthenelse{\boolean{dc}}
{}
{\clearpage}

\begin{appendix}[D]
\section*{\begin{center}Determining the S3T stability of the homogeneous state to non-zonal perturbations\end{center}}\label{app:nzstability}

The S3T stability   of  a homogeneous background state to non-zonal perturbations of the form $e^{\sigma t} e^{\i(mx+ny)}$, in which $m$ is the zonal wavenumber of the perturbations and $n$ is the meridional wavenumber of perturbations, is determined by  the eigenvalues $\sigma$ that solve for each $(m,n)$  the equation:
\ifthenelse{\boolean{dc}}
{
\begin{align}
& \s+ \rU \d_{m0} + r(1-\d_{m0}) -\i m \beta/N^2 = \nonumber\\
&~=\sum_{k}\sum_{\ell} 2(n k - m \ell)\[\bit  m n (k^2_+-\l^2_+)-(m^2-n^2)k_+\l_+\]\times\nonumber\\
&~~~~~~~~\times(K^2/N^2-1) \frac{\varepsilon\,\tilde{Q}_{k\l}}{2r}  {\bigg /} \[\bit (\s+2r)K^4K^2_s+\right.\nonumber\\
&~~~~~~~~\left.+2\i\beta k_+(k_+m+\l_+n)K^2-\i m \beta K^2 (K^2+K^2_s)/2\bit \]~.\label{eq:s3tgenstab}
\end{align}
}
{
\begin{align}
& \s+ \rU \d_{m0} + r(1-\d_{m0}) -\i m \beta/N^2 = \nonumber\\
&~~~=\sum_{k}\sum_{\ell}  \frac{2(n k - m \ell)\[ m n (k^2_+-\l^2_+)-(m^2-n^2)k_+\l_+\](K^2/N^2-1) }{(\s+2r)K^4K^2_s+2\i\beta k_+(k_+m+\l_+n)K^2-\i m \beta K^2 (K^2+K^2_s)/2}\frac{\varepsilon\,\tilde{Q}_{k\l}}{2r} ~.\label{eq:s3tgenstab}
\end{align}
}
In the above equation, $N^2=m^2+n^2$, $K^2=k^2+\l^2$, $K^2_s=(k+m)^2+(\l+n)^2$, $k_+=k+m/2$, $\l_+=\l+n/2$, $\delta_{ij}$ is Kronecker's delta and $\tilde{Q}_{k\l}$ are the Fourier coefficients of the forcing covariance as defined in Eq.~\eqref{eq:Q_ab_to_Q_kl_posneg}. 
This form of the equation is appropriate for a square domain of length $2 \pi$ and the summations are over the integer 
wavenumbers $k$ and $\l$. 
The derivation of the above equation can be found in \cite{Bakas-Ioannou-2013-jfm}. For a specified forcing with spectrum $\tilde{Q}_{k\l}$  the growth rates are obtained using Newton's method.


At high supercriticality, i.e. as  $\varepsilon \to \infty$, the maximal growth rate, $\s_r$, of the $(m,n)$ large scale structure asymptotes to:
\be
\s_r^2  =  \frac{\varepsilon}{r}\,   \Phi \(m,n, \tilde{Q} \)~,
\ee
with 
\ifthenelse{\boolean{dc}}
{
\begin{align}
\Phi&\(m,n, \tilde{Q} \)= \nonumber\\
&=\sum_{k}\sum_{\ell} 
\[\bit  m n (k^2_+-\l^2_+)-(m^2-n^2)k_+ \l_+\]\times\nonumber\\
& \qquad\qquad\qquad  \times (n k - m \ell) \(K^2/N^2-1\) \frac{\tilde{Q}_{k\l}}{  K^4K^2_s}~,
\end{align}
}
{
\begin{align}
\Phi\(m,n, \tilde{Q} \)=\sum_{k}\sum_{\ell} (n k - m \ell) \[\bit  m n (k^2_+-\l^2_+)-(m^2-n^2)k_+ \l_+\] \(K^2/N^2-1\) \frac{\tilde{Q}_{k\l}}{  K^4K^2_s}~,
\end{align}
}
and the frequency of this eigenstructure assumes asymptotically  the Rossby wave frequency, $\s_i= \imag{(\s)}=m\beta/( m^2 + n^2)$. This asymptotic expression for the growth rate and phase speed of the large scale structure is useful for tracing the maximal growth rates as a function of supercriticality  using Newton's iterations.  

\begin{figure}[h!]
\centering\includegraphics[width=19pc,trim=3mm 10mm 3mm 5mm,clip]{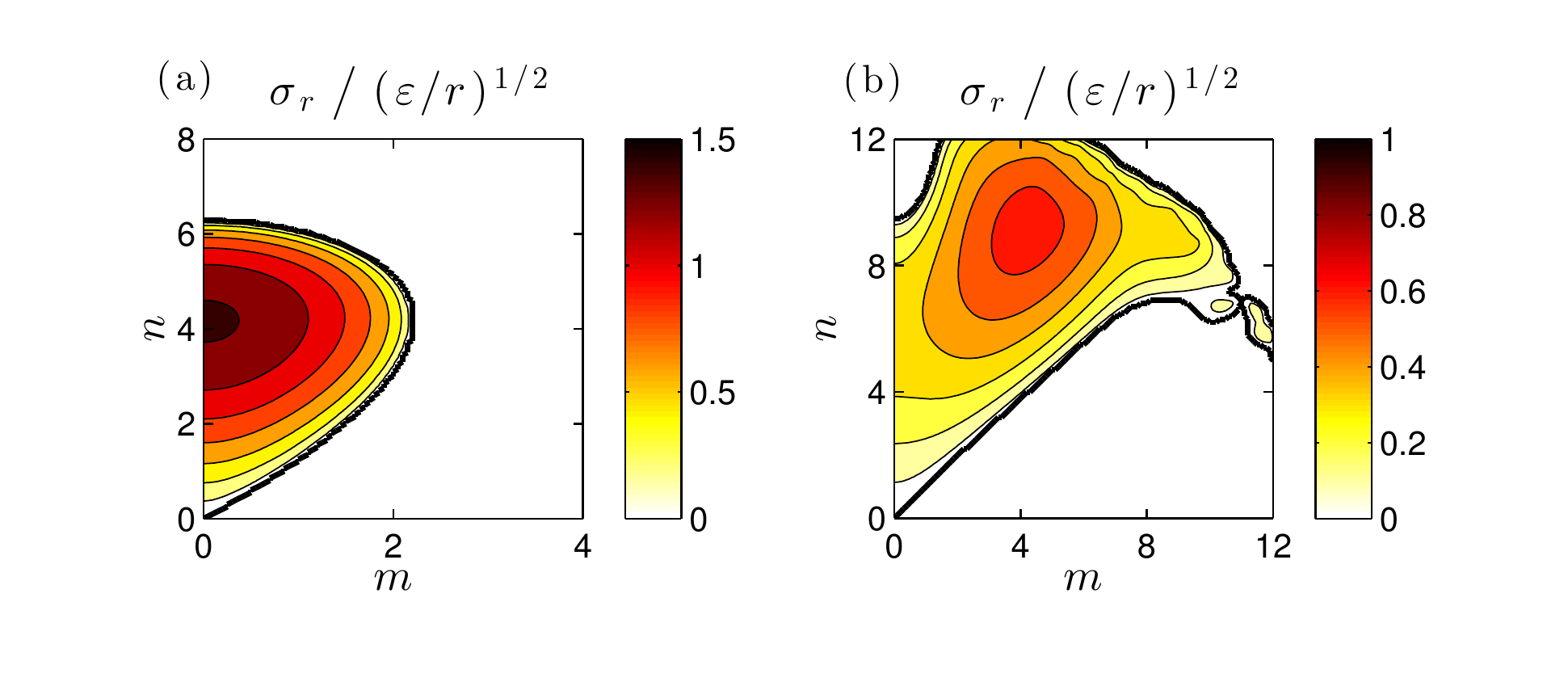}
\caption{The $\varepsilon\to+\infty$ asymptotic maximal growth  rate $\s_r$ scaled by $\sqrt{\varepsilon/r}$ as a function of the wavenumbers  $(m,n)$ of the S3T eigenfunction. In this asymptotic limit  $\sigma_r \ge 0$, and the zero contour is marked with a thick solid line. The asymptotic growth rate is independent of $\beta$ and  depends only the forcing spectrum. Shown are the asymptotic growth rates for NIF (panel (a)) and IRFn (panel (b)). In NIF maximal instability occurs for jet structures, while in IRFn maximal instability occurs for non-zonal structures.}
  \label{fig:asymptotic}
\end{figure}

The asymptotic growth rates  depend only on the forcing distribution. The growth rates  for the NIF and IRFn  forcing used in this paper are shown in  Fig.~\ref{fig:asymptotic}.  It can be shown that the asymptotic growth rate vanishes for exactly isotropic forcing. 
Asymptotically the growth rates do not depend on the damping rate of the mean flow, $\rU$.  NIF forcing favors at least initially jet formation, while IRFn favors  formation of non-zonal structures.

\end{appendix}

\end{document}